\documentclass[fleqn,usenatbib]{mnras}

\usepackage{newtxtext}
\usepackage{mathptmx}

\usepackage[T1]{fontenc}

\usepackage{hyperref}
\usepackage{graphicx}
\usepackage{txfonts}
\usepackage{lipsum}
\usepackage{subcaption}         
\usepackage{lscape}             
\usepackage{placeins}           
                                
\DeclareRobustCommand{\VAN}[3]{#2}
\let\VANthebibliography\thebibliography
\def\thebibliography{\DeclareRobustCommand{\VAN}[3]{##3}\VANthebibliography}

\usepackage{graphicx}	

\title[A cell model for methane biosignatures]{A generalised microbial cell model for methane biosignature predictions}

\author[A.E. Nicholson \& N.J. Mayne]{
Arwen. E. Nicholson$^{1}$\thanks{E-mail: arwen.e.nicholson@gmail.com},
Nathan J. Mayne$^{1}$
\\
$^{1}$Department of Physics and Astronomy, University of Exeter
}

\date{Accepted XXX. Received YYY; in original form ZZZ}

\pubyear{\the\year{}}

\begin{document}
\label{firstpage}
\pagerange{\pageref{firstpage}--\pageref{lastpage}}
\maketitle

\begin{abstract}
The majority of potentially habitable planets detected to date are likely quite different to Earth, for example, being larger in radius and mass, differing rotation rates and with host star spectra unlike the Sun. Therefore the first alien life detected will potentially be living in conditions not found on our planet. This necessitates a generalised approach to modelling biology that can be applied to numerous planetary scenarios, built on fundamental knowledge of life on Earth, but not limited by it.
Here, we explore a generalised model of a microbial cell, whose metabolic rate is governed by thermodynamics and substrate diffusion across its cell wall. We model a single-species biosphere consisting of methane producing microbes and determine how changing the cell size, cell death rate and biomass synthesis cost influence the biosignature on the planet - in this case methane. We discuss approaches to predicting upper estimates for the biosignature gas abundance and the applicability of the model to other metabolisms. This tool adds to the body of work attempting to grapple with the complexity of potential alien biospheres.
\end{abstract}




\section{Introduction}
\label{section:introduction}

Numerous potentially habitable planets have been identified from observational data \citep{hill2023catalog}, however constraining the surface conditions of these exoplanets, let alone whether they host life, is enormously challenging \citep{montet2015stellar, foreman2015systematic, cloutier2019confirmation,tsiaras2019water, madhusudhan2020interior, wogan2024hycean, shorttle2024distinguishing}.
Developing our abiotic models of planetary atmospheres is of course crucial for understanding the observed data from any exoplanet. However life has the potential to strongly influence the atmosphere of its planet, for example acting to `erase' certain abiotic features by transforming abiotically produced gases via their metabolisms \citep{lovelock1965physical, lovelock1974atmospheric}. Therefore, models including biology are key when considering potentially inhabited planets, not only for the identification of biosignatures, but also for understanding the wider planetary context of any observational data \citep{Catling:2018, Meadows:2018, Krissansen-Totton:2022,arthur2025edge}.

Given the range of planets discovered, it is likely that many potentially habitable planets experience very different surface conditions to those of Earth. For example, the most abundant planets in our galaxy have no solar-system analogue; these super-Earth and sub-Neptune planets have sizes and masses between those of their namesake solar system planets \citep{zhu201830,livingston2026young}. A proposed subset of these planets are Hycean planets, which are characterised by global oceans and rich $H_{2}$ atmospheres, and are thought to be potentially habitable \citep{madhusudhan2021habitability, madhusudhan2023chemical}. Hycean planets would be significantly easier to observe than rocky Earth-like planets due to their extended atmosphere and so are of great interest for potential biosignature observations.
Observational limitations also mean that the smaller rocky planets that have been detected to date generally have orbits that are close in to host stars that are significantly smaller and cooler than the Sun \citep{sterzik2012biosignatures,fujii2018exoplanet,quanz2022,Snellen:2021}. Therefore, most potentially habitable exoplanets detected to date likely experience very different surface conditions than those on Earth due to factors such as the size and mass of the exoplanet, the orbital radius, and the stellar class of the host star. In our search for life in the galaxy, we are therefore likely searching for alien life that lives in a very different environment to any Earth-based life. This necessitates determining generalised properties of life that we can reasonably expect to find in alien biospheres. Towards this aim we propose a generalised biological model of methane producing microbes that can be used within a wide diversity of planetary contexts to understand a potential methane biosignature. The abiotic part of the model in this work is based on conditions found on early-Earth in a simplified manner. This part of the model acts as a testing-ground for our biological experiments so as to determine the qualitative impact on the planetary environment by changing various microbe parameters. To form predictions for a particular exoplanet a specific set up for the exoplanet in question, taking into account many factors such as planet size, mass, host star etc would be required as these can all impact how a biologically produced gas might accumulate in the planet's atmosphere \citep{segura2005biosignatures, kiang2018exoplanet}. 

In models of pre-Cambrian Earth it is standard to model microbial life by fixing the growth yield ratio of a metabolism (the energy required to build biomass) and the concentration down to which life can draw oceanic nutrients. These parameters, and a chosen burial rate, i.e. the biomass buried on the ocean floor and not recycled by heterotrophic life, determine the accumulation of by-products in the environment \citep{kharecha2005coupled, bruggeman:2014}. We build on this understanding to build a generalised cell model that can be used in different planetary contexts to inform biosignature prediction models.

In this work, we consider properties of Earth-based life that we assume to be universal across the galaxy. We restrict this work to considering microbial chemosynthetic life and aim to expand this approach in later work to consider photosynthetic microbial and plant life. We assume that alien-life would utilise chemical potential gradients in their environment and obtain energy following the Gibbs free energy equation. We also assume that uptake of nutrients to an alien cell will be limited by diffusion uptake in some way. We consider free-living microbes, within an ocean, that obtain nutrients via diffusion across their cell membranes. While many lifeforms on Earth can move to seek out food they are still restricted by diffusion across cell membranes for their critical life processes e.g. uptake of $CO_{2}$ through plant leaves or $O_{2}$ diffusion across lung membranes in many animals. The ability to evolve and adapt is also typically included in definitions of life \citep{vitas2019towards} and therefore we assume competition for resources will occur within any alien biospheres. 

To develop our model of a generalised cell model for a methane producing chemosynthetic microbial biosphere we use a quasi-realistic representation of the early-Earth. Forming biosignature predictions requires both plausible biological models and detailed models of the planetary environment. The same microbial life would result in different biosignatures under different planetary contexts.
Here our focus is on the biological side of this puzzle and so as our planetary model is roughly based on early-Earth, the model as a whole cannot be used as-is for forming biosignature predictions. Any biosignature predictions for a potentially habitable planet will require bespoke modelling of the planetary environment. 
Numerous planetary parameters play a role in influencing the model behaviour, including the planet's mass and atmospheric pressure. The depth, coverage and dynamics of the planet's ocean will also influence the atmosphere. We aim to explore the influence of these parameters on possible biosignatures in future work. 
Hycean planets are of particular interest with regards to potential biosignature observations as their large size makes for easier observations compared to those of Earth-sized planets. Adding the biotic component of the model presented in this work into an abiotic Hycean model is another avenue of future work we hope to explore.

We focus here on methane since this potential biosignature can be readily detectable using current telescope technology \citep{thompson2022case, bell2023methane, wogan2024hycean, lunine2025characterization}. Methane has been detected in the atmospheres of K2-18b \citep{bezard2022methane, madhusudhan2023carbon, fernandez2026atmospheric}, and TOI-270d \citep{benneke2024jwst, holmberg2024possible}, both candidate Hycean exoplanets situated within the habitable zone of their stars. However, the detection of methane cannot automatically be taken as a biosignature as abundant methane is also detected in similarly-sized planets that are not thought to be habitable, such as the sub-Neptune planet LP791-18c which is not thought to have a solid surface \citep{roy2024jwst}. For terrestrial planets a detection of abundant methane would potentially constitute a stronger sign of life as there are no known abiotic pathways for rocky planets to acquire high quantities of methane \citep{wogan2020abundant, seager2025prospects}. However with their smaller size these planets are harder to observe and currently only those with close orbits to M-dwarf stars are accessible for atmospheric studies \citep{trappist2024roadmap}. So far there has been no definitive detections of an atmosphere around a terrestrial planet \citep{lunine2025characterization}. Both the robust detections of methane in the atmospheres of possible Hycean planets, and the potential strength of methane as a biosignature for terrestrial planets motivates building models of methane producing biospheres that are minimally bound by assumptions based on Earth-based life.

In this work we aim to produce a generalised microbial cell model that can be added to abiotic models of exoplanets to determine the maximum accumulation of methane gas in the observable regions of the atmosphere of the planet as a by-product of life, given an assumed rate of $H_{2}$ and $CO_{2}$ outgassing. This will help guide observational searches by determining whether a predicted biosignature would accumulate in quantities high enough to be observable. The microbial cell model can also be used to calculate the necessary supply of $H_{2}$ to the biosphere required to yield a certain abundance of methane for a specific exoplanet. By predicting the abiotic source of gases to a biosphere we can validate whether the predicted abundance of abiotic gases agrees with other measured properties of the planet, e.g. a predicted high $H_{2}$ outgassing rate on a planet would imply a high rate of volcanic activity, which would leave other evidence in a planet's atmosphere \cite{edmonds2022volcanic}.

Any microbe cell model will unavoidably be specific to the given metabolism, however the approach to building the model cell taken here can be applied to nutrient limited chemosynthetic microbe cells with other metabolisms. How the biosignature accumulates in the environment will be specific to the metabolism itself and the planetary context \citep{thompson2022case}. Therefore this works presents a generalised microbial cell model for methane biosignature predictions, but the wider approach can be adopted for prediction of biosignature due to other nutrient limited chemosynthetic life.

Previous work by \cite{nicholson2022} showed that for a simple nutrient-limited chemosynthetic microbial biosphere, the availability of the limiting substrate to the biosphere was significantly more important in determining the resulting biosignature gas in the atmosphere, than the specific biological parameters describing the microbial life, e.g. cell death rates and energy maintenance costs. Diffusion-limited substrate uptake was not included in \cite{nicholson2022}, and microbes in this model were able to consume nutrients down to concentrations of zero. Here we extend the model to include diffusion-limited substrate uptake with the aim of a more realistic and generalised microbe model that is more readily adaptable to different planetary scenarios. We find that including diffusion-limitation impacts the biosphere's ability to draw down $H_{2}$ when the biotic parameters describing the microbes change and thus in turn impacts the biosignature. 
By arguing that life will evolve to exploit limiting nutrients, and using studies determining the bounds of life as we know it, we propose a generalised microbial model for methane biosignature prediction modelling. 

This paper is structured as follows. Section \ref{section:methods} outlines the model description and the parameter space explored in this work. We base our microbe model on lab measurements grounding it in known realistic biology, with model steps follow realistic microbe behaviours. In Section \ref{section:results} we explore the impact of including diffusion-limited $H_{2}$ uptake on the atmosphere biosignature. In Section \ref{section:plausible_cell} we argue that competition and evolution are likely to occur within alien biospheres allowing us to form predictions that limit the biological parameter space required to explore when formulating biosignature predictions for a chemosynthetic microbial biosphere.
Finally, in Section \ref{section:conclusions} we present our conclusions.

\section{Methods}
\label{section:methods}

In this Section we briefly cover the planetary setup used in the following simulations and the biological parameters used to describe the model microbes. The physical parameters and planetary setup follow those used in \cite{nicholson2022} and so a more complete description can be found there. The code used for the following simulations is also publicly available on github: \url{https://github.com/nicholsonae/biosignatures}.

\subsection{Planet setup}

\begin{figure*}
\centering
\includegraphics[scale=0.33]{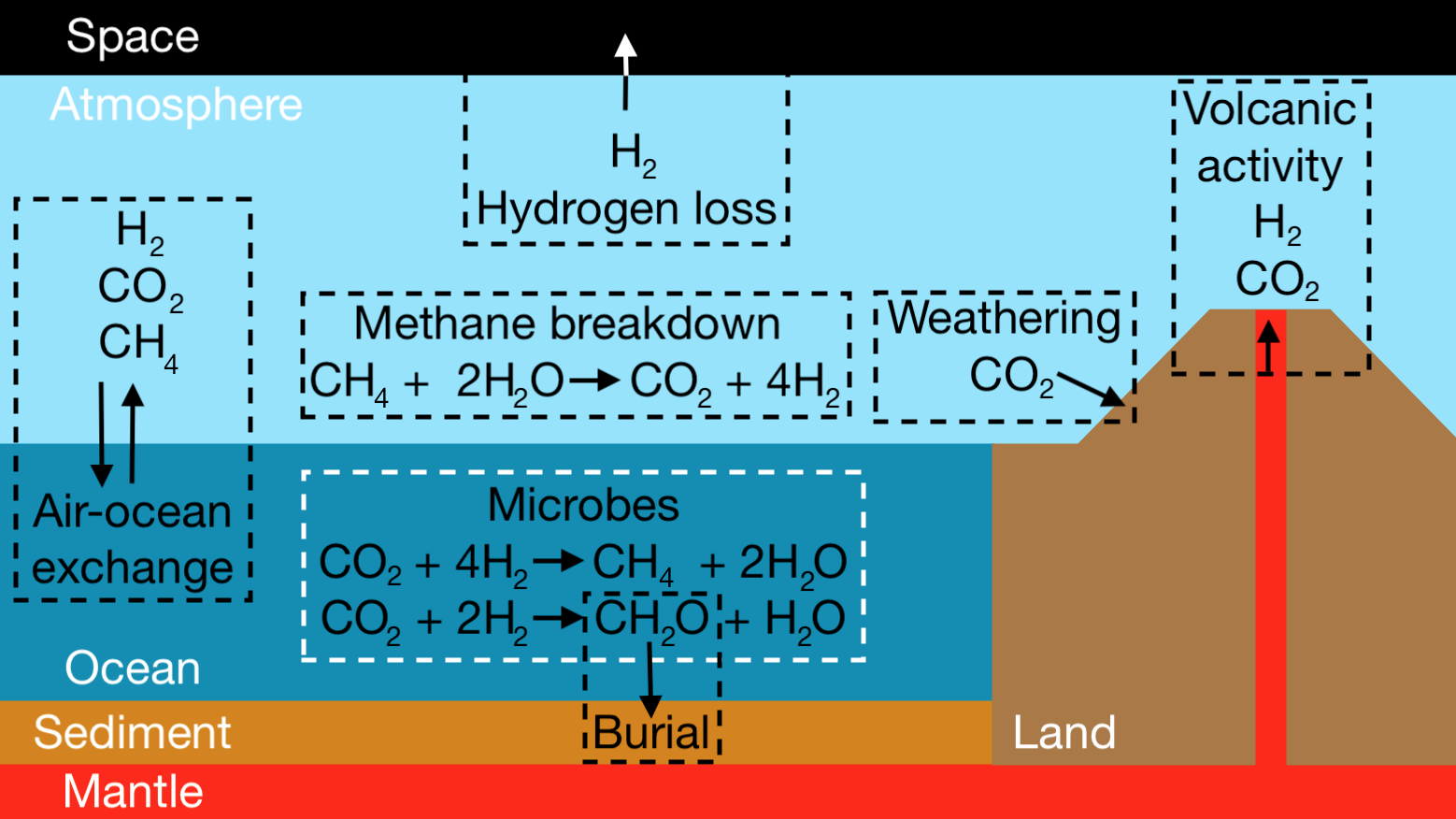}
\caption{A schematic showing the key abitoic (black dashed boxes) and biotic (white dashed box) processes occurring in the model \citep[reproduced from][]{nicholson2022}.}
\label{fig:schematic}
\end{figure*}

For our planet setup we follow \cite{nicholson2022} and assume an Earth-sized planet covered in a global ocean that orbits a sun-like star. As in \cite{nicholson2022} the model planet's atmosphere and ocean are represented in 0D. We assume an $N_{2}$ dominated atmosphere and abundant $H_{2}O$ for all reactions that require it. We track the movement of $CO_{2}$, $H_{2}$, and $CH_{4}$ through the system.  
We assume a constant atmospheric pressure, $P_{atmo} = 1\ atm$, and a constant total number of moles of gas comprising the atmosphere $n_{atmo} = 1.73\times 10^{20}\ mol$ taking modern Earth values. 
The abiotic environment is updated with the inflows / losses of each traced gas in model time-steps representing years.
Figure \ref{fig:schematic} shows a schematic of the model set up.

\subsubsection{Atmosphere setup}

We assume constant sources of $H_{2}$, and $CO_{2}$ to the atmosphere in an approximation of volcanic outgassing. On Earth, $H_{2}$ is lost from the upper atmosphere layers irreversibly to space via diffusion of $H_{2}$. Assuming a dry stratosphere \citep{Hunten:1973, Walker:1977}, the rate of hydrogen loss is proportional to $f(H_{2}) + 2f(CH_{4})$ where $f(H_{2})$ and $2f(CH_{4})$ are the mixing ratios of $H_{2}$ and $CH_{4}$ respectively. We assume a constant rate of atmospheric hydrogen loss according to these ratios.
In our model atmospheric $CO_{2}$ removal occurs by removing a fixed percentage of the atmospheric $CO_{2}$ each year. This acts as a highly simplified representation of silicate weathering. Abiotic $CO_{2}$ removal from Earth's atmosphere is far more complex than represented here \citep{west2005tectonic}, and the dependence of seafloor weathering on Earth's climate is currently less understood \citep{brady1997seafloor} and this process can act as either a sink or a source of $CO_{2}$ to the ocean, adding an additional layer of complexity. As we are primarily interested in the biotic component of our system a simplified removal of $CO_{2}$ is suitable.

We include no abiotic source of $CH_{4}$ and on an Earth-like planet with no atmospheric $O_{2}$, $CH_{4}$ is long lived in the atmosphere and its breakdown occurs via photolysis at the top of the atmosphere. This is a complex process that is altitude dependant and has several steps, and so we \textcolor{blue}{simplify} it here for our 0-D representation of a planet's atmosphere and assume that methane breaks down according to $CH_{4} + 2H_{2}O \rightarrow CO_{2} + 4H_{2}$ following \citet{kharecha2005coupled} and \citet{nicholson2022}. We assume that this $CH_{4}$ breakdown occurs at a fixed rate proportional to the quantity of methane in the atmosphere.

The surface temperature of our model planet is $CO_{2}$ and $CH_{4}$ dependant. The parametrisation of temperature against atmospheric composition was generated using the Met Office Unified Model (UM) \citep{Boutle:2017, Eager:2020} through simulated snapshots of the planet's temperature for different levels of $CO_{2}$ and $CH_{4}$ in the atmosphere. To generate these snapshots the UM is set up as described in \citet{eager20233d}, assuming a 1 billion year old host star of type G and a planet radius of $6051.3\times 10^{3}m$. Atmospheric configurations are then run to equilibrium for different atmospheric quantities of $CH_{4}$ and $CO_{2}$ and these snapshots are then interpolated into a 2D grid which informs the model planet's temperature based on the atmospheric composition. For full details see \citet{nicholson2022, eager20233d}.

\begin{table}
 \caption[]{Parameters for the abiotic influx and outflux of atmospheric $CO_{2}$, $H_{2}$, and $CH_{4}$, where $T(X)$ is the total number of moles of molecule $X$ in the atmosphere.}
 \begin{center}
    
\begin{tabular}{ccc}
 \hline \hline
 \\
 Chemical & Influx ($yr^{-1}$) & Outflux ($yr^{-1}$) \\
 \\ \hline 
$CO_{2}$   & $10^{15}$ & $0.001\times T(CO_{2})$ \\
$H_{2}$ & $10^{13}$ &  $0.001\times (T(H_{2}) + 2T(CH_{4}))$ \\
$CH_{4}$  & $0$ & $0.001\times T(CH_{4})$\\
\hline \hline

\end{tabular}
 \end{center}
\label{table:abiotic}
\end{table}

Table \ref{table:abiotic} shows the abiotic influxes and outfluxes of $H_{2}$, $CO_{2}$, and $CH_{4}$ to the planet atmosphere used for all of the following simulations. Abiotic influxes are kept fixed throughout simulations. We model the outgassing and abiotic removal of $H_{2}$ and $CO_{2}$ in an idealised way with values chosen to ensure a habitable surface suitable for seeding with microbes. To form biosignature predictions for a particular planet these parameters would need to be bespoke for the planet in question as factors such as the mass of the planet, and the properties of the host star will affect the rate of $H_{2}$ loss from a planet's atmosphere \citep{hunten1976hydrogen, miller2009atmospheric, savanov2021activity}.

\subsubsection{Ocean-atmosphere gas exchange}
\label{subsection:ocean_atmo_gas_exchange}

We assume that the only source of $CO_{2}$ and $H_{2}$ to the ocean is from the atmosphere. Gases in the atmosphere dissolve into the ocean and become available to the biosphere, and biotically produced methane (the only source of methane to the system) is exchanged between the ocean and the atmosphere. 
We calculate the molecular transfer of gases between the ocean and atmosphere following the stagnant boundary layer model \citep{liss1974flux} where the exchange of gases depends on the concentration gradient through a thin film of water on the top of the ocean (full details can be found in Section \ref{section:appendix_gas_exchange}).

\subsection{Microbe model}
\label{subsection:microbe_model}

We model a single-species single-celled chemosynthetic microbial biosphere, based on the Methanosarcina barkeri species, where microbes generate energy from the metabolism via
\begin{equation}
\label{eq:metabolism}
    CO_{2} + 4H_{2} \rightarrow CH_{4} + 2H_{2}O,
\end{equation}
where the energy obtained from this reaction is given by the Gibbs free energy expressed as
\begin{equation}
\label{eq:gibbs}
    \Delta G = \Delta G^{0} + RT\ log(Q).
\end{equation}
$\Delta G^{0}$ is the free energy change of the reaction under standard conditions and we use $\Delta G^{0} = (-253 + 0.41\ T)kJ\ mol^{-1}$ following \cite{Kral:1998}, $R$ is the universal gas constant ($0.008314\ kJ\ mol^{-1}\ K^{-1}$), $T$ is the temperature in kelvin.
For $\Delta G_{CH_{4}}$ - the energy required to synthesise 1 $mol$ of $CH_{4}$ - $Q$ is given by
\begin{equation}
\label{eq:gibbs_q}
    Q = \frac{[CH_{4}]^{*}_{aq} \cdot a(H_{2}O)^{2}}{[CO_{2}]^{*}_{aq} \cdot ([H_{2}]_{aq}^{*})^{4}}.
\end{equation}
where $[i]^{*}_{aq}=\frac{[i]_{aq}}{\alpha(i)}$ is the dissolved concentration of species $i$ divided by its Henry's law coefficient (its solubility). The values of $\alpha(H_{2})$, $\alpha(CO_{2})$, and $\alpha(CH_{4})$ can be found in Table \ref{table:gas_air_exchange} and $a(H_{2}O)$ is assumed to be 1 \citep{kharecha2005coupled}.
We will explore two scenarios, one where $\Delta G$ is given by Equation \ref{eq:gibbs} and another where we fix $\Delta G$ to be a constant value.

Our model microbes build their biomass following
\begin{equation}
\label{eq:biomass_building}
    CO_{2} + 2H_{2}\ (+ATP) \rightarrow CH_{2}O + H_{2}O.
\end{equation}
We use $CH_{2}O$ as a proxy for the microbes' biomass. We base the default microbe parameters around those determined in lab experiments for Methanosarcina barkeri, which gives us a baseline for modelling a realistic life-form. We can then change various parameters describing the microbes to test how changing these parameters change the microbes' impact on their environment.

In \citet{nicholson2022} we included a parameter for the ATP maintenance cost of the microbes. Here we remove this parameter and instead vary the energy cost for biosynthesis. We can represent the energetic cost of microbes equivalently by either including a separate ATP maintenance cost, or by having a higher biosynthesis cost. Provided the overall energetic cost per mole of biomass is the same, whether the energy a microbe generates is for maintenance costs or biosynthesis costs is irrelevant and will produce the same amount of $CH_{4}$. See Appendix \ref{section:appendix_A0} for further discussion.

\begin{table*}
 \caption[]{Default biological parameters used for our model microbes. These values come from lab based measurements of the Methanosarcina barkeri, a single-celled methane producing archea. Values labelled $*$ are taken from \citep{Lynch:2019}. Values labelled $\dagger$ are based on results from \citet{servais1985}.}
 \begin{center}
    
\begin{tabular}{ccc}
 \hline \hline
 \\
 Cell parameter & Default values & Sensitivity tests \\
 \\ \hline 
radius  & $r_{0}\ =\ 10^{-6}\ m$ $^{*}$ &  $0.1 r_{0}$, $0.5 r_{0}$, $0.75 r_{0}$, $r_{0}$, $1.25 r_{0}$, $1.5 r_{0}$\\
death rate  & $d_{0}\ =\ 0.02\ h^{-1}\ ^{\dagger}$ & $0.0$, $0.005$, $0.02$, $0.04$, $0.07$, $0.1$ \\
biomass density  & $b_{0}\ =\ 3530$ $mol_{CH_{2}O}/m^{3}$ $^{*}$ &   \\
$\Delta G_{CH_{2}O}$ & $m_{0} = 97.5\ kJ\ mol^{-1}_{CH_{2}O} $ $^{*}$ & $0.5m_{0}, m_{0}, 1.5m_{0}, 2m_{0}, 3m_{0}, 5m_{0}, 10m_{0}, 20m_{0}$\\
\hline \hline

\end{tabular}
 \end{center}
\label{table:microbes}
\end{table*}

The microbe cell parameters that we vary are the cell radius (we assume cells to be spherical), the rate at which cells die, and the energy cost of synthesising 1 mole of biomass. We also have to include the energy cost to build biomass and generate ATP. We fix the biomass density of cells. We calculate the default values of these parameters based on data from lab experiments on microbes. Overall microbe cell mortality rates have been measured to be between $0.01 - 0.03\ h^{-1}$ in aquatic environments \citep{servais1985} and so we use $0.02\ h^{-1}$ as our default value here. A full list of the microbe parameters can be found in Table \ref{table:microbes}. Some of the parameters are taken directly from lab measurements and some we have adapted to use in our model. For full details on how the default values in Table \ref{table:microbes} are calculated, see Appendix \ref{section:appendix_A1}

\subsubsection{Diffusion-limited substrate uptake}
\label{section:diff_info}

At steady state the flux of a substrate $F$ {($mol\ s^{-1}$)}, through the surface of a spherical of cell radius $r$ ($m$), is given by:
\begin{equation}
\label{eq:diff_limited}
    F = 4 \pi r D S_{\infty},
\end{equation}
where $D$ is the diffusion coefficient ($m^{2}s^{-1}$) for the substrate and $S_{\infty}$ is the concentration of the substrate ($mol\ m^{-3}$) far from the cell wall \citep{Berg1977}. Equation \ref{eq:diff_limited} makes the assumption that the cell is a perfect sink for the substrate molecules, i.e. a cell completely absorbs all molecules that reach its surface. Equation \ref{eq:diff_limited} therefore is a theoretical maximum diffusion-limited substrate uptake rate.

\cite{ARMSTRONG20081311} follow \cite{pasciak1974transport, pasciak1975transport} in deriving the theoretical maximum diffusion-limited substrate uptake flux across a cell membrane. This uptake rate, $F_{D}(r)$, depends on distance from the centre of the cell cell, $r$, and is given by
\begin{equation}
F(r) = 4 \pi D r^{2} \frac{\partial S}{\partial r},
\end{equation}
where $S_{r}$ is the substrate concentration at a distance $r$ from the centre of the cell.
\begin{equation}
    \int^{S_{\infty}}_{S_{o}} \partial S(r) = \frac{F(r_{0})}{4 \pi D} \int^{\infty}_{r_{0}} r^{-2}\partial r,
\end{equation}
equates to
\begin{equation}
[S_{\infty} - S_{0}] = \frac{F(r_{0})}{4 \pi D} r_{0}^{-1},
\end{equation}
which can be rearranged to give
\begin{equation}
F(r_{0}) = 4 \pi D r_{0} [S_{\infty} - S_{0}].
\end{equation}
We then make the assumption that $S_{0} = 0$, i.e. the cell absorbs all molecules of the substrate that reach its surface. Thus, we reach Equation \ref{eq:diff_limited} as our theoretical upper limit to diffusion limited substrate uptake by our microbes. While this is an upper limit, our model neglects other strategies microbes have evolved to uptake substrates such as moving to areas of higher substrate concentration and changing body shape to maximise surface area against the volume ratios, (e.g. growing appendages or movement \citep{pavlova2022bacterial}). We choose here to model a spherical cell to simplify both modelling diffusion into the cell and to enable changing the cell size with a single parameter - its radius. The effects of non-spherical cell shapes can be included in Equation \ref{eq:diff_limited} by multiplying by a dimensionless shape coefficient \citep[see][]{ARMSTRONG20081311,pasciak1975transport, pahlow1997impact}. As our aim here is to consider a qualitative study of an idealised and general model, and not to generate specific quantitative predictions, we focus on spherical cells. Including the impact of motion involves the more complex Sherwood number \citep[see][]{ARMSTRONG20081311, karp1996nutrient}, which we retain for future work. As we are using an Archean Earth-like setup, $H_{2}$ is the limiting substrate to the microbes and so their growth is limited by its availability. We use a diffusion coefficient for $H_{2}$ of $D_{H_{2}} \approx 4.3 \times 10^{-9} m^{2}s^{-1}$ \citep{wang2023diffusion} to calculate the maximum $H_{2}$ uptake of a microbe, and then the $CO_{2}$ uptake is calculated depending on how the $H_{2}$ is `allocated' by the microbe to either energy generation or biomass generation (Section \ref{Section:h2_allocation}). Note that for a different type of planet, e.g. a Hycean planet with a hydrogen-dominated atmosphere, $CO_{2}$ would instead be the limiting substrate to microbe growth, and so the same approach could be taken but for determining the $CO_{2}$ availability to the microbes instead of the $H_{2}$ availability.

\subsubsection{Calculating cell $H_{2}$ allocation}
\label{Section:h2_allocation}

Microbes must obtain energy via Equation \ref{eq:metabolism} which utilises $H_{2}$, and also use $H_{2}$ to build biomass following Equation \ref{eq:biomass_building}. The ratio at which $H_{2}$ is ``assigned'' to either biomass or energy generation depends on the energy obtained from the microbes' metabolism, Equation \ref{eq:metabolism}. The energy generated per 1 mole of $CH_{4}$ metabolised via Equation \ref{eq:metabolism} is given by $\Delta G_{CH_{4}}$, Equation \ref{eq:gibbs}. Therefore the energy obtained per mole of $H_{2}$ is given by
\begin{equation}
    \frac{\Delta G_{CH_{4}}}{4}.
\end{equation}
We use values from \citet{Lynch:2019} for the energy required to synthesise 1 mole of $ATP$, and the number of moles of $ATP$ required to build 1 mole of $CH_{2}O$ to calculate the total energy required to build 1 mole of $CH_{2}O$ (see Appendix \ref{section:appendix_A1} for further details).

The energy required to metabolise 2 moles of $H_{2}$ into 1 mole of biomass ($CH_{2}O$) is given by
\begin{equation}
\Delta G_{ATP} \times ATP_{CH_{2}O} = \Delta G_{CH_{2}O}
\end{equation}
where $\Delta G_{ATP}$ is the energy required to synthesise a mole of $ATP$ and $ATP_{CH_{2}O}$ is the number of moles of $ATP$ required to synthesise 1 mole of $CH_{2}O$. Therefore the number of moles of $H_{2}$ required for energy generation in order to build 1 mole of $CH_{2}O$ is 
\begin{equation}
   \frac{4\ \Delta G_{CH_{2}O}}{\Delta G_{CH_{4}}}
\end{equation}
As we require 2 moles of $H_{2}$ to build 1 mole of $CH_{2}O$ then the number of moles of $H_{2}$ required for energy generation per mole of $H_{2}$ used for biomass synthesis is given by
\begin{equation}
\label{eq:ratio}
  H_{2}^{biomass}\ :\ H_{2}^{energy}\ =\ 1\ :\ \frac{2\ \Delta G_{CH_{2}O}}{\Delta G_{CH_{4}}}.
\end{equation}
The maximum $H_{2}$ uptake by a cell is given by the diffusion-limited rate (Equation \ref{eq:diff_limited}), and microbes will then assign $H_{2}$ to either energy generation or biomass synthesis according Equation \ref{eq:ratio}.

\subsection{Model structure}
\label{subsection:model_structure}

Microbial growth rates are often given in terms of units of an hour \citep[e.g.][]{Weissman:2021}, whereas climate simulations tend to stabilise over the course of years or decades to changes in atmospheric composition \citep{eager20233d}, therefore the abiotic environment is updated on timesteps representing years and the microbe biosphere is updated on timesteps representing hours. The microbe population is recorded in terms of the total biomass of the biosphere, and $ATP$ available to the biosphere. The population of the biosphere can be calculated by dividing the total biomass of the biosphere by the biomass contained within one microbe cell. 

Once the model is initialised, it is run for $20,000$ years before seeding with life to allow the atmosphere and ocean to reach equilibrium. Life is then seeded at $t = 20,000$ years and the model is run for a further $20,000$ years to allow the system to reach a steady state after the introduction of life (which happens by roughly 10,000 years after the appearance of life (see Figure \ref{fig:0_b}). The atmospheric and oceanic abundances of $H_{2}$, $CO_{2}$, and $CH_{4}$ are recorded along with the microbe population. See Appendix \ref{section:appendix_A3} for further details on the timesteps of the model.
To test the sensitivity of the biological parameters on our results we change one biological parameter at a time, keeping the others fixed to the default ones listed in the `Default values' column of Table \ref{table:microbes}. 

\section{Results}
\label{section:results}

\begin{figure*}
\centering
\begin{subfigure}{.45\textwidth}
  \centering
  \includegraphics[scale=0.47]{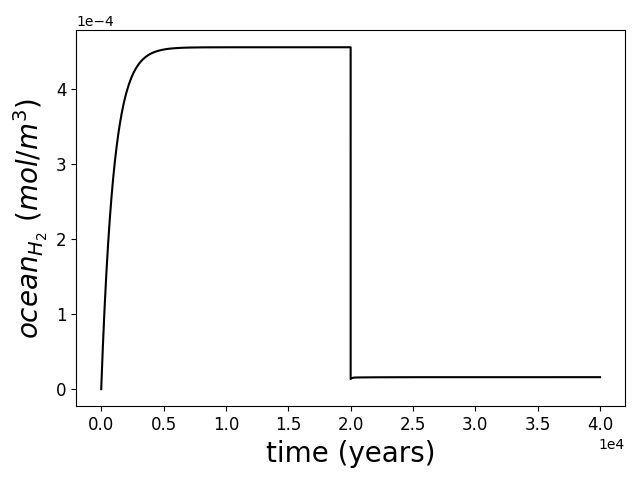}
  \caption{Concentration of $H_{2}$ in the ocean over time.}
  \label{fig:0_a}
\end{subfigure}
\hfill
\begin{subfigure}{.45\textwidth}
  \centering
  \includegraphics[scale=0.47]{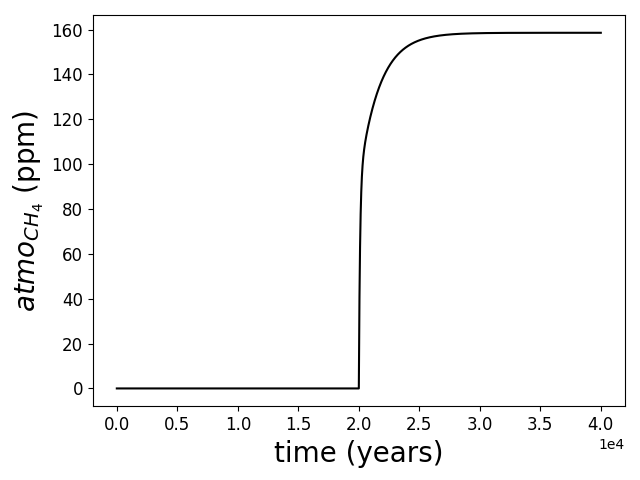}
  \caption{Level of $CH_{4}$ in the atmosphere against time.}
  \label{fig:0_b}
\end{subfigure}%

\centering
\caption{Panels showing the concentration of $H_{2}$ in the ocean, $ocean_{H_{2}}$ and the level of methane in the atmosphere $atmo_{CH_{4}}$ over time. Life is introduced to the system at $t = 20,000$ years.}
\label{fig:0}
\end{figure*}

When life is introduced to the planet the concentration of $H_{2}$ in the ocean rapidly drops before stabilising at a low concentration due to the microbes' consumption of $H_{2}$, and the level of $CH_{4}$ in the atmosphere rises rapidly. 
Figure \ref{fig:0_a} and \ref{fig:0_b} shows the concentration of $H_{2}$ in the ocean, and the abundance of $CH_{4}$ in the atmosphere, respectively, over time. In each case, the introduction of life at $t = 20,000$ years is clearly seen. The concentration at which ocean $H_{2}$ stabilises depends on the rate at which microbes require to uptake $H_{2}$ to maintain a stable population. As described in Section \ref{section:diff_info} the flow of molecules into the cell depends on the concentration of the substrate in the ocean. A stable microbial population is only achieved when the birth rate (which depends on the uptake rate of $H_{2}$ by the microbe) matches the death rate of the microbes (which is a set parameter, see Table \ref{table:microbes}). Therefore the concentration of $H_{2}$ in the ocean stabilises at a `limiting concentration'. Below this limiting concentration the population of the microbes will drop, leading to fewer microbes removing $H_{2}$ from the ocean and thus allowing the concentration of $H_{2}$ to rise via diffusion from the atmosphere, and above the limiting concentration the population of the microbes will increase and thus $H_{2}$ will be removed more rapidly from the ocean lowering its concentration. In this way the concentration of $H_{2}$ in the ocean is regulated by the biosphere \citep[see][for further details]{nicholson2022}.

The rest of this section presents results on how changing the cell parameters of cell radius, cell death rate, and the energy cost of biomass synthesis impact the biotic $CH_{4}$ output for scenarios where cell uptake of $H_{2}$ is limited by diffusion, as given by Equation \ref{eq:diff_limited}. In Section \ref{subsection:changing_radius_death_rate} we discuss the impact of changing the cell radius or cell death rate, and then in Section \ref{subsection:changing_ch2o_cost} we discuss how changing the energetic cost of biomass ($CH_{2}O$) synthesis changes the model dynamics.

\subsection{Changing the cell radius or cell death rate}
\label{subsection:changing_radius_death_rate}

\begin{figure*}
\centering
\begin{subfigure}{.45\textwidth}
  \centering
  \includegraphics[scale=0.47]{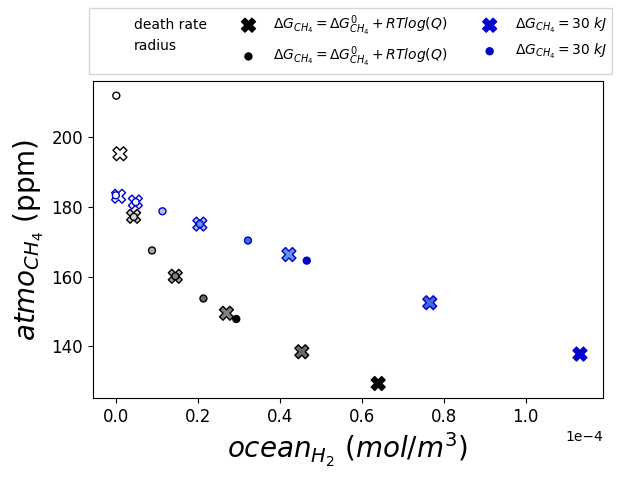}
  \caption{Level of $CH_{4}$ in the atmosphere against the concentration of $H_{2}$ in the ocean.}
  \label{fig:1_a}
\end{subfigure}
\hfill
\begin{subfigure}{.45\textwidth}
  \centering
  \includegraphics[scale=0.47]{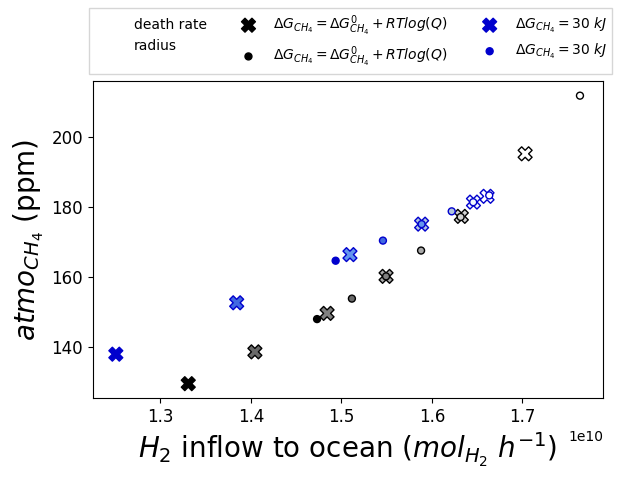}
  \caption{Level of $CH_{4}$ in the atmosphere against the the rate of $H_{2}$ inflow into the ocean.}
  \label{fig:1_b}
\end{subfigure}%

\centering
\caption{Panels showing data from experiments with different cell death rates, marked with an $\times$, and different cell radii, marked with an $\circ$, for experiments where $\Delta_{CH_{4}}$ is given by Equation \ref{eq:gibbs}, shown in black, and for experiments with fixed $\Delta_{CH_{4}} = 30\ kJ$, shown in blue. Marker colour saturation indicates the parameter value for the $CH_{2}O$ synthesis cost with a darker marker indicating a higher cost of $CH_{2}O$ synthesis.}
\label{fig:1}
\end{figure*}

Figure \ref{fig:1_a} shows level of methane in the atmosphere, $atmo_{CH_{4}}$ against the concentration of hydrogen in the ocean $ocean_{H_{2}}$ for simulations with differing cell death rates (shown with $\times$ markers) and radii (shown with $\circ$ markers) as listed in Table \ref{table:microbes}. There are two energy scenarios shown, those where $\Delta G_{CH_{4}}$ is given by Equation \ref{eq:gibbs}, shown in black, and data where $\Delta G_{CH_{4}} = 30\ kJ$, shown in blue. The colour saturation of each point correlates with the parameter value, i.e. a darker filled circle indicates a large cell radius, and a lighter filled circle indicates a small cell radius. Each data-point represents a microbial biosphere with fixed cell parameters throughout the simulation, and for experiments of different cell death rates, the cell radius is set to the default value in Table \ref{table:microbes}, and similarly for experiments of different cell radii, the cell death rate is set to the default value in Table \ref{table:microbes}.

Figure \ref{fig:1_a} shows an overall trend in the data that as we increase either the cell size, or the cell death rate (indicated by a higher marker colour saturation) the level of $H_{2}$ in the ocean increases, and the level of methane in the atmosphere decreases. The correlation between $atmo_{CH_{4}}$ and $ocean_{H_{2}}$ is linear where $\Delta G_{CH_{4}}$ is fixed (blue data), and non linear when $\Delta G_{CH_{4}}$ is given by Equation \ref{eq:gibbs}. Figure \ref{fig:1_b} shows the same data but this time for the level of methane in the atmosphere against the rate of hydrogen inflow to the ocean. In Figure \ref{fig:1_b} we can see that $atmo_{CH_{4}}$ increases linearly with $H_{2}$ inflow to the ocean when $\Delta G_{CH_{4}}$ is fixed (blue data). This agrees with previous findings of \cite{nicholson2022}, i.e. that the strength of a biosignature linearly increases with an increase of the availability of the limiting nutrient to the biosphere. The inflow of $H_{2}$ to the ocean scales with the concentration of $H_{2}$ in the ocean (transfer of $H_{2}$ from the atmosphere to the ocean increases as $ocean_{H_{2}}$ decreases, see Appendix \ref{section:appendix_gas_exchange} for more details).

\begin{figure*}
\centering
\begin{subfigure}{.45\textwidth}
  \centering
  \includegraphics[scale=0.47]{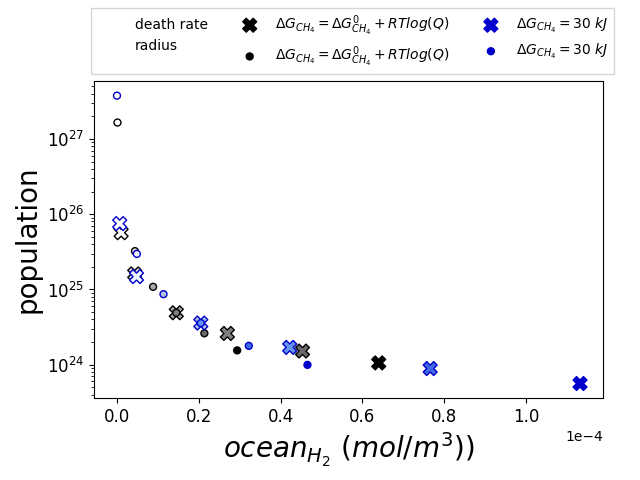}
  \caption{Total population of the biosphere as a function of the concentration of $H_{2}$ in the ocean.}
  \label{fig:2_a}
\end{subfigure}
\hfill
\begin{subfigure}{.45\textwidth}
  \centering
  \includegraphics[scale=0.47]{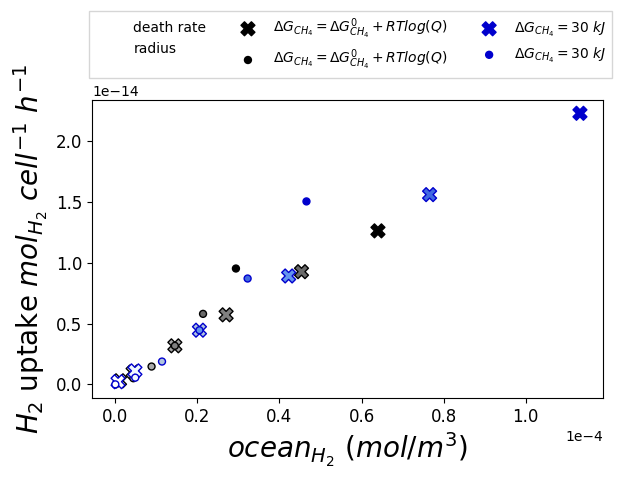}
  \caption{Rate of $H_{2}$ uptake per microbe cell as a function of the concentration of $H_{2}$ in the ocean.}
  \label{fig:2_b}
\end{subfigure}%

\centering
\begin{subfigure}{.45\textwidth}
  \centering
  \includegraphics[scale=0.47]{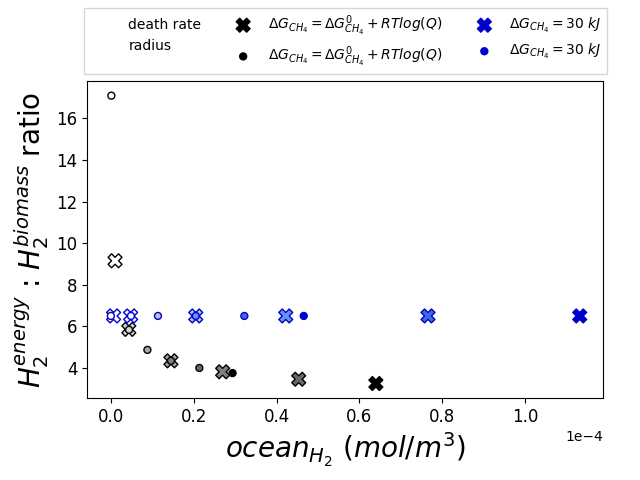}
  \caption{Ratio of $H_{2}$ allocated to energy generation as a function of the biomass synthesis against the concentration of $H_{2}$ in the ocean.}
  \label{fig:2_c}
\end{subfigure}
\hfill
\begin{subfigure}{.45\textwidth}
  \centering
  \includegraphics[scale=0.47]{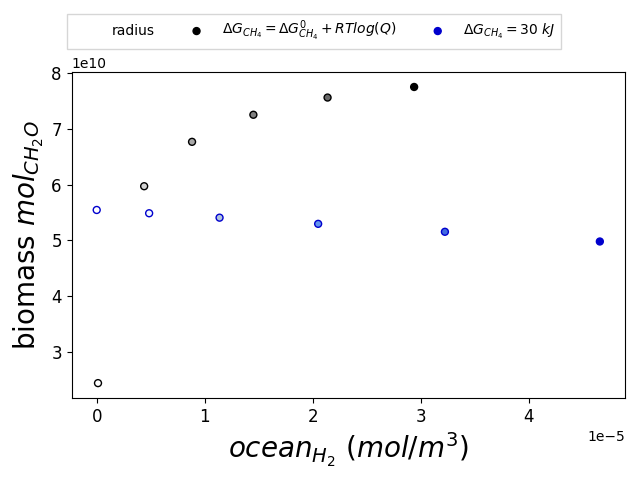}
  \caption{Total biomass of biosphere as a function of the concentration of $H_{2}$ in the ocean.}
  \label{fig:2_d}
\end{subfigure}%

\centering
\caption{Panels showing data from experiments with different cell death rates, marked with an $\times$, and different cell radii, marked with an $\circ$, for experiments where $\Delta_{CH_{4}}$ is given by Equation \ref{eq:gibbs}, shown in black, and for experiments with fixed $\Delta_{CH_{4}} = 30\ kJ$, shown in blue. Marker colour saturation indicates the parameter value for the $CH_{2}O$ synthesis cost with a darker marker indicating a higher cost of $CH_{2}O$ synthesis.}
\label{fig:2}
\end{figure*}

For a stable population the birth rate of cells must match the death rate, meaning cells must accumulate sufficient biomass via Equation \ref{eq:biomass_building} to build a new cell before dying. Given the same cell death rate, a smaller cell will need to build biomass at a slower rate to achieve this compared to a larger cell. This leads to smaller cells having lower $H_{2}$ requirements per hour than larger cells.  The $H_{2}$ cell uptake rate depends on $ocean_{H_{2}}$ (see Equation \ref{eq:diff_limited}) with faster $H_{2}$ uptake rates occurring for higher values of $ocean_{H_{2}}$. Therefore $ocean_{H_{2}}$ will be maintained at a lower level for a biosphere consisting of smaller cells due to the lower uptake rate of $H_{2}$  that is required for a stable population. If $ocean_{H_{2}}$ drops below a level that can sustain a stable population, microbes will begin to starve and thereby as fewer microbes are removing $H_{2}$ from the ocean, the $ocean_{H_{2}}$ increases. If $ocean_{H_{2}}$ is higher than the level required for a stable population, microbes will reproduce a rate above the cell death rate, leading to a higher population and thus a reduction in $ocean_{H_{2}}$. This feedback leads to biotic regulation of the concentration of $H_{2}$ in the ocean. Reducing the death rate of cells corresponds to a lower $ocean_{H_{2}}$ for a similar reason. Longer lived cells do not need to build biomass as rapidly as shorter lived cells and so have a lower $H_{2}$ uptake requirement and thus $ocean_{H_{2}}$ is maintained at a lower level. 

The level of methane in the atmosphere depends both on the availability of hydrogen to the biosphere, and the energy available to the biosphere when synthesising $H_{2}$ to $CH_{4}$ which in turn depends on the concentrations of $H_{2}$, $CH_{4}$ and $CO_{2}$ in the ocean, see Equations \ref{eq:gibbs} and \ref{eq:gibbs_q}. Figure \ref{fig:2} shows the population of the biosphere (\ref{fig:2_a}), the $H_{2}$ uptake rate per microbe cell (\ref{fig:2_b}), the $H_{2}$ allocation to either energy generation or biomass (\ref{fig:2_c}), and the biomass of the biosphere (\ref{fig:2_d}) against the the concentration of $H_{2}$ in the ocean for both cases with a fixed $\Delta G_{CH_{4}}$ (shown in blue), and where $\Delta G_{CH_{4}}$ varies as given by Equation \ref{eq:gibbs}. As before the colour saturation of each point correlates with the parameter value, with darker saturations indicating higher parameter values. From Figures \ref{fig:2_a} and \ref{fig:2_b} we find that increasing either the cell death rate or cell size leads to a decrease in the biosphere's population, and increase in individual cell $H_{2}$ uptake (as well as the increase in $ocean_{H_{2}}$ already discussed). Figures \ref{fig:1_a} \ref{fig:1_b} show that an increase in $ocean_{H_{2}}$ correlates with a decrease in the inflow of $H_{2}$ from the atmosphere to the ocean, thus decreasing the availability of $H_{2}$ to the biosphere. The individual microbe's $H_{2}$ uptake needs set the concentration of $H_{2}$ in the ocean as discussed above and this uptake rate increases linearly with $ocean_{H_{2}}$ as we increase the cell death rate, and increases at a faster rate when increasing the cell size. The radius of the cell also impacts the $H_{2}$ uptake rate (see Equation \ref{eq:diff_limited}), with faster uptake for higher cell radii, causing this difference in behaviour. Therefore increasing cell death rate or cell size reduces the total availability of $H_{2}$ to the biosphere leading to a decrease in the level of methane in the atmosphere. 

Figure \ref{fig:2_c} shows the ratio of $H_{2}$ used for energy generation as opposed to biomass building as a function of ocean $H_{2}$. When $\Delta G_{CH_{4}}$ is fixed (blue data) this ratio is constant and this leads to the linear relationship between $atmo_{CH_{4}}$ and $ocean_{H_{2}}$ / $H_{2}$ inflow to the ocean seen in Figure \ref{fig:1}. However when $\Delta G_{CH_{4}}$ is given by Equation \ref{eq:gibbs}, it increases as $ocean_{H_{2}}$ increases and thus the ratio of $H_{2}$ going to energy against the biomass decreases with increasing $ocean_{H_{2}}$. As the microbes can achieve more energy per mole of $CH_{4}$ synthesised they can more efficiently use $H_{2}$ to build biomass. For very low concentrations of $H_{2}$ in the ocean the energy yield for $CH_{4}$ synthesis is low and so microbes must use more $H_{2}$ for energy synthesis which increases the biosignature. 

Figure \ref{fig:2_d} shows the biomass of the biosphere as a function of $ocean_{H_{2}}$ for the cell radius experiments. When changing the cell death rate the biomass within a cell is unchanged and so the biomass scales with population (Figure \ref{fig:2_a}. However, when we change the cell radius, the relationship between the population and the biomass of the biosphere changes. Figure \ref{fig:2_d} shows that where $\Delta G_{CH_{4}}$ is fixed the biomass drops as we increase the cell size (indicated by darker circles) and $ocean_{H_{2}}$ decreases albeit slower than the population declines. The increased cell size leads to the biomass of the biosphere declining much more slowly than the population declines when $\Delta G_{CH_{4}}$ is fixed. However, when $\Delta G_{CH_{4}}$ is given by Equation \ref{eq:gibbs}, as the cell size increases, the biomass of the biosphere increases despite the population decreasing. The increased energy obtained by microbes per mole of $CH_{4}$ synthesised allows for more $H_{2}$ to be used for biomass synthesis and thus a lower biosignature and a higher amount of biomass than for fixed $\Delta G_{CH_{4}}$ scenarios.

\subsection{Changing the biomass synthesis cost}
\label{subsection:changing_ch2o_cost}

\begin{figure*}
\centering
\begin{subfigure}{.45\textwidth}
  \centering
  \includegraphics[scale=0.47]{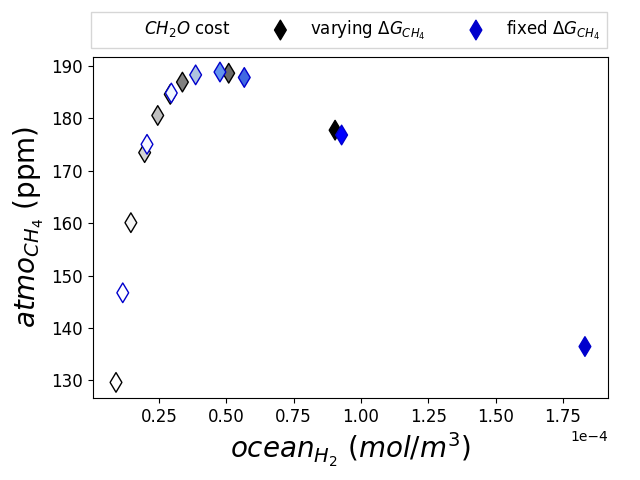}
  \caption{Level of $CH_{4}$ in the atmosphere as a function of the concentration of $H_{2}$ in the ocean.}
  \label{fig:3_a}
\end{subfigure}
\hfill
\begin{subfigure}{.45\textwidth}
  \centering
  \includegraphics[scale=0.47]{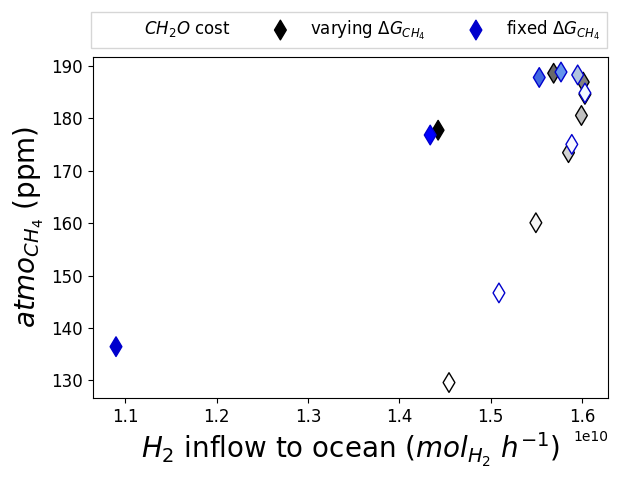}
  \caption{Level of $CH_{4}$ in the atmosphere as a function of the rate of $H_{2}$ inflow into the ocean.}
  \label{fig:3_b}
\end{subfigure}%

\centering
\begin{subfigure}{.45\textwidth}
  \centering
  \includegraphics[scale=0.47]{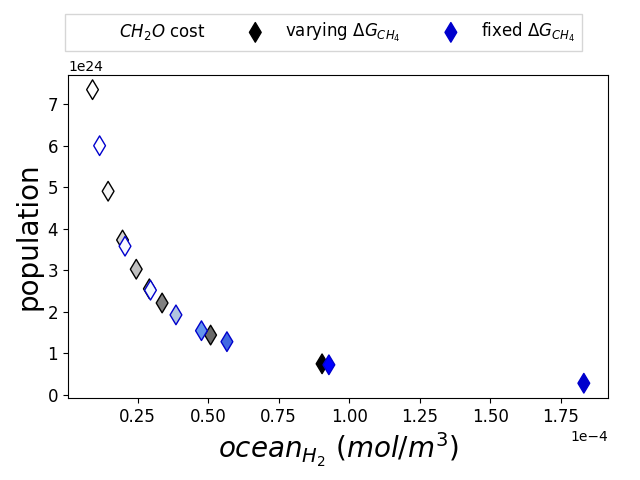}
  \caption{Total population of the biosphere as a function of the concentration of $H_{2}$ in the ocean.}
  \label{fig:3_c}
\end{subfigure}
\hfill
\begin{subfigure}{.45\textwidth}
  \centering
  \includegraphics[scale=0.47]{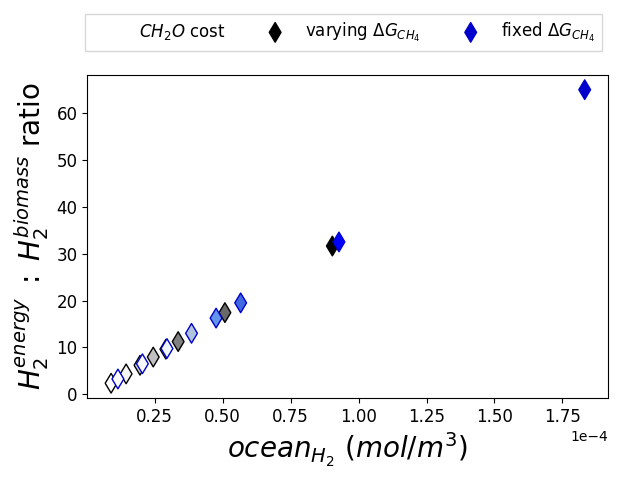}
  \caption{Ratio of $H_{2}$ allocated to energy generation as a function of the biomass synthesis against the concentration of $H_{2}$ in the ocean.}
  \label{fig:3_d}
\end{subfigure}%

\centering
\caption{Panels showing data from experiments with different $CH_{2}O$ synthesis costs for experiments where $\Delta_{CH_{4}}$ is given by Equation \ref{eq:gibbs}, shown in black, and for fixed $\Delta_{CH_{4}} = 30\ kJ$, shown in blue. Marker colour saturation indicates the parameter value for the $CH_{2}O$ synthesis cost with a darker marker indicating a higher cost of $CH_{2}O$ synthesis.}
\label{fig:3}
\end{figure*}

Changing the energy cost of synthesising 1 mole of $CH_{2}O$ impacts the model behaviour differently to changing the cell size or cell death rate. Figure \ref{fig:3_a} shows $atmo_{CH_{4}}$ against $ocean_{H_{2}}$ for a range of experiments with different $CH_{2}O$ synthesis costs (see Table \ref{table:microbes}). As before darker marker saturation correlates with a higher $CH_{2}O$ cost. Figure \ref{fig:3_a} shows an initial increase in $atmo_{CH_{4}}$ with increasing $ocean_{H_{2}}$ as the $CH_{2}O$ cost increases, before decreasing after a peak. Figure \ref{fig:3_b} shows $atmo_{CH_{4}}$ against the inflow of $H_{2}$ to the ocean. We see again a peak in $atmo_{CH_{4}}$ however as we increase the $CH_{2}O$ cost (indicated by darker markers) we see initially that $atmo_{CH_{4}}$ increases with an increased inflow of $H_{2}$ to the ocean, but that after the $atmo_{CH_{4}}$ peak the $H_{2}$ inflow to the ocean begins to decrease with increasing $CH_{2}O$ cost. This shows that there are two distinct $atmo_{CH_{4}}$ scenarios that can result from the same $H_{2}$ inflow rate to the ocean, this is due to the recycling of methane in the atmosphere, as described in Section \ref{section:methods} which impacts the level of $H_{2}$ in the atmosphere which also determines the rate of inflow of $H_{2}$ to the ocean. This leads to the same rate of $H_{2}$ inflow to the ocean being possible for two different concentrations of $H_{2}$ in the ocean. See Appendix \ref{section:Appendix_A4} for experiments where methane recycling in the atmosphere is omitted.

Figure \ref{fig:3_c} shows the biosphere population as a function of $ocean_{H_{2}}$ for changing $CH_{2}O$ synthesis costs. As the biomass contained within a cell is constant throughout the $CH_{2}O$ synthesis cost experiments the biomass of the biosphere will scale with the population. As with increasing either the cell radius or cell death rate, Figure \ref{fig:3_c} shows that increasing the $CH_{2}O$ cost decreases the population while $ocean_{H_{2}}$ increases. The cell death rate is the same for all experiments and so therefore is the reproduction rate, however for microbes with higher $CH_{2}O$ costs, microbes require a faster rate of $H_{2}$ uptake to maintain a stable population and therefore require higher concentrations of $H_{2}$ in the ocean. With more $H_{2}$ needed for energy generation as the cost of $CH_{2}O$ synthesis increases only lower populations can be supported by the environment. Figure \ref{fig:3_d} shows the ratio of $H_{2}$ used for energy generation as a function of biomass synthesis and shows a linear relationship with increasing $ocean_{H_{2}}$. The $H_{2}$ uptake of each microbe cell also scales linearly with $ocean_{H_{2}}$ (see the death rate data in Figure \ref{fig:3_d}). Initially, the drop in population does not offset the increased $CH_{4}$ output per microbe due to the increasing energy costs of synthesising biomass, and so $atmo_{CH_{4}}$ initially increases. At $ocean_{H_{2}} \approx 0.5\ mol/m^{3}$ the balance between the drop in population and the increased $CH_{4}$ output per microbe changes and $atmo_{CH_{4}}$ begins to decrease with increasing $ocean_{H_{2}}$ as the cost of $CH_{2}O$ is further increased.

\section{Formulating plausible biosignature predictions}
\label{section:plausible_cell}

In Section \ref{section:results} we analysed how changing the biological parameters governing the behaviour of model microbes, namely cell death rate, cell size, and biomass energy synthesis costs, impacts the the total population of the biosphere, the quantity of biologically produced $CH_{4}$, and the concentration of $H_{2}$ maintained in the ocean by the biosphere. 
Figure \ref{fig:1_a} shows that increasing any of the cell parameters of cell size, cell death rate or the energy cost of biomass synthesis act to increase the concentration of $H_{2}$ in the ocean, as increasing these parameters necessitates a faster rate of $H_{2}$ uptake per microbe for the biosphere to maintain a steady population. On Earth competition for resources is a crucial driver of adaptation and evolution and most definitions of life include the ability to adapt via natural selection \citep{vitas2019towards}, therefore it is a process we expect to be present in an alien biosphere - the organism most able to exploit a limiting resource will outcompete other species / subspecies and come to dominate that `niche' in the environment. Within our model this competition would lead to the microbe able to drawn $H_{2}$ in the ocean down to the lowest concentrations outcompeting species than can only function at higher $H_{2}$ concentrations in the ocean due to needing faster rates of $H_{2}$ uptake. Therefore from our results, microbes that are smaller, longer lived, and with lower energy biomass synthesis costs will outcompete those which are shorter lived, larger, more dense, or have higher energy costs. 
For the parameters of cell size and cell death rate, smaller values lead not only to lower concentrations of $H_{2}$ in the atmosphere but also higher levels of $CH_{4}$ being produced by the biosphere and thus accumulating in the atmosphere. Therefore a microbial biosphere of longer lived, smaller cells would not only outcompete larger, shorter lived cells, but would also be easier for us to observe remotely. 
We can't have an infinitely small cell, as life requires some minimum size and biomass to contain genetic information, a cell wall, energy storage, etc. 
Definitions of life also exclude it from being immortal, or at least the cells consisting of the organism from being immortal. Entropy dictates that cells will be required to spend energy on maintaining / replacing biomass.
The smallest cells on Earth are heterotrophs, however when considering biosignatures we want an approximation for the primary producers of any alien biosphere. 
The smallest cell sizes for methanogenic life have been measured as having a length of $0.6\mu m$ and width of $0.1\mu m$ \citep{michal2018phymet2} which would correspond to a radius of $\approx 0.2\mu m$ for a spherical cell assuming the same volume.  
\cite{ortega2024minimum} investigates minimal energy costs for cell building and present a `minimal' cell \citep{hutchison2016design, breuer2019essential, moger2023evolution}. This minimal cell is able to replicate and fulfil all its metabolic needs and is smaller than any known free-living cell on Earth. As briefly discussed in Section \ref{section:diff_info} the effects of shape and motion on the uptake of nutrients by cells can be explored by including the relevant parameters to Equation \ref{eq:diff_limited} \citep[see][]{ARMSTRONG20081311}. These kinds of studies can be used to determine a reasonable estimate for the minimum size of a hypothetical alien cell to use in biosignature prediction models. 

A lower bound for the death rate of microbe cells is hard to constrain even for microbes on Earth \citep{wu2024microbes} as microbes can survive for months during periods of nutrient or water scarcity \citep{hobbie2013microbes, leung2020energetic, mason2022microbial}, or for thousands of years in a dormant state while frozen in ice \citep{caro2025frozen}. However microbes in these latter scenarios only metabolise very slowly as they spend much of their time in stasis and so would be unlikely to produce observable biosignatures. Any life likely to leave an observable biosignature will be in an active, metabolising state. Measurements of `residual' microbe cell death (e.g. cell death not due to predation) in marine ecosystems has been measured as being between $0.002 - 0.02\ h^{-1}$ \citep{servais1985}. In a multi-species ecosystem grazing behaviour by heterotrophs may become important and additional mortality from grazing has been measured to be $0 - 0.02\ h^{-1}$ for marine ecosystems \citep{servais1985}.

The measured energy cost for biomass synthesis (which is typically recorded in units of moles of ATP required per mole of biomass in laboratory experiments) for microbes varies significantly.
The work of \citet{ortega2024minimum} aims to calculate a minimal energy cost for building a cell and includes a more realistic description of the components of a cell than presented here (where cells consist simply of `biomass' represented by $CH_{2}O$). Across 4 different cell types they find a very similar minimal energy cost of cell synthesis per gram which gives confidence that generalised models of cells can be applied in astrobiological settings. They note that there is little available thermodynamic data available for biopolymers and utilise group contribution methods to estimate energetic costs for biomass synthesis. Here, for methane biosignatures, we have identified that there is a peak in the $CH_{4}$ produced by a biosphere when changing the biomass synthesis cost. For our goal of determining the maximum biosignature plausibly produced by a biosphere, we can identify where this peak $CH_{4}$ output occurs for the microbe cells under the conditions predicted to be present on the exoplanet in question as this peak depends both on the parameters chosen for the other biological parameters describing the microbes' cells, and on the physical environment.

The model discussed in this work can be used in two ways in astrobiology research. Firstly, given a modelled abiotic planet with estimations for abiotic parameters such as $H_{2}$ outgassing, the generalised model can be used to form a maximum biosignature strength by inputting the plausible minimal estimations for cell size, a low cell death rate, and the energetic cost of biomass synthesis where the biosignature peaks (see Figure \ref{fig:3_a}).
Alternatively the model can be used to calculate abiotic parameters of a planet where there has been a tentative detection of life. On a planet with life, life will act to `erase' abiotic aspects of their environment. For example in the models discussed in this work, despite a high rate of $H_{2}$ being supplied to the environment, $H_{2}$ levels in the atmosphere and oceans are kept much lower due to life consuming $H_{2}$. If a planet with abundant methane was detected, this model can be used to calculate the minimum $H_{2}$ influx required for a simple methanogen biosphere to produce the biosignature. This could help yield a prediction for e.g. how volcanically active the planet might be, and this could be compared with other observable evidence to corroborate or falsify this prediction \citep{edmonds2022volcanic}.

While not the focus of this work, this model can also yield a prediction for the amount of biomass present to produce a potentially observed biosignature. This enables us to link predictions from our model with the biomass based model of determining biosignature plausibility developed by \citet{Seager_2013} which use data from life on Earth to estimate whether the biomass required to produce a proposed biosignature is plausible.

\section{Conclusions}
\label{section:conclusions}

In this work we have demonstrated that a simplified model of chemosynthetic microbial life, whose growth is limited by substrate availability, can be represented by a few key assumptions the cell metabolism, the cell size, the rate of cell death, and the energetic cost of biomass synthesis. We make the argument that we would expect evolution and competition for resources to be a feature of alien biospheres (and these processes are required for most definitions of life) and so we would expect alien life to evolve to exploit their environment to some limiting degree. In this model the limiting factor on microbe growth rate is $H_{2}$ availability. We find that decreasing cell death rates, cell sizes and cell biomass densities all lead to lower concentrations of $H_{2}$ in the ocean and also higher abundances of $CH_{4}$ in the atmosphere, which in our model acts as our biosignature. Therefore, due to resource competition we would except alien life to evolve to be smaller and longer lived. We can use studies investigating the smallest cell possible to inform our model. Constraining the cell death rate is less straightforward as e.g. the presence of high levels of UV radiation, or extreme temperatures will impact the rate of cell damage and death, as will the presence or absence of secondary consumers. When forming biosignature predictions, we can use measurements of microbe cell death rates both with and without the affects of grazing to inform our models e.g. \cite{servais1985}.  %

We would expect real alien biospheres to be vastly more complex than our simple model here, not only because alien biospheres are likely to consist of multiple species \citep{arthur2017entropic, arthur2022selection, arthur2023dice}, some of whom consume the outputs of others, which will dramatically impact the biosignature. For example, secondary consumers that recycle the biomass of dead methane producing microbes increasing the methane signature in the atmosphere. The aim of this work is to provide a baseline model for a microbial cell, where different metabolisms can be modelled and predictions for alien biospheres can start to be built where the model assumptions are clearly laid out and minimal in number. 

Earth's biosphere is dominated by plant and microbial life \citep{bar2018biomass} and these lifeforms are essential for more complex life such as mammals to evolve. Earth's microbial world has been likened to the `life support system' of our biosphere due to the fundamental role it plays in shaping the surface of our planet \citep{cavicchioli2019scientists}. The majority of Earth's remotely observable biosignatures \citep{sterzik2012biosignatures} are the products of either plant life, or microbial life, such as the `red-edge', a sharp increase in albedo in wavelengths over $700nm$ due to surface vegetation, and a large disequilibrium in the atmospheric abundances of methane and oxygen. Therefore constraining our possible parameter space for describing hypothetical alien plant or microbe life is vital for forming biosignature predictions for a range of alien biosphere scenarios. This does assume that an alien biosphere would similarly have a fundamental foundation of microbial life supporting any larger lifeforms as is the case for Earth's biosphere. In this work we haven taken steps to develop a minimal cell model that can be used for biosignature predictions for nutrient-limited chemosynthetic life, and in future work plan to extend this approach to photosynthetic microbes and plants. 

The goal of detecting and ultimately verifying the presence of life on a planet other than Earth requires extensive, multi-disciplinary effort, and simple, idealised, models informed by life on Earth, but minimally constrained by assumptions relevant to the specifics of Earth life, are required as part of this effort. Much more work is required, both in terms of increasing model complexity, exploring the impact of additional processes and feedbacks, for example, dynamics/circulations and atmospheric chemistry, and simplified approaches exploring the primary mechanisms ultimately controlling the observational signatures.

\section*{Acknowledgements}
This work was supported by a UKRI Future Leaders Fellowship and extension [grant numbers MR/T040866/1 $\&$ MR/Z000122/1]. This work was partly funded by the Leverhulme Trust through a research project grant [RPG-2020-82]. We'd like to thank the two anonymous reviewers for their helpful feedback in improving this manuscript.

\section*{Data Availability}
The code used to generated the data in this study can be found at: \url{https://github.com/nicholsonae/biosignatures}



\bibliographystyle{mnras}
\bibliography{main} 

@ARTICLE{wogan2024hycean,
       author = {{Wogan}, Nicholas F. and {Batalha}, Natasha E. and {Zahnle}, Kevin J. and {Krissansen-Totton}, Joshua and {Tsai}, Shang-Min and {Hu}, Renyu},
        title = "{JWST Observations of K2-18b Can Be Explained by a Gas-rich Mini-Neptune with No Habitable Surface}",
      journal = {\apjl},
         year = 2024,
        month = mar,
       volume = {963},
       number = {1},
          eid = {L7},
        pages = {L7},
          doi = {10.3847/2041-8213/ad2616},
archivePrefix = {arXiv},
       eprint = {2401.11082},
 primaryClass = {astro-ph.EP},
       adsurl = {https://ui.adsabs.harvard.edu/abs/2024ApJ...963L...7W}
}

@article{lovelock1974atmospheric,
  title={Atmospheric homeostasis by and for the biosphere: the Gaia hypothesis},
  author={Lovelock, James E and Margulis, Lynn},
  journal={Tellus},
  volume={26},
  number={1-2},
  pages={2--10},
  year={1974},
  publisher={Taylor \& Francis}
}

@article{arthur2017entropic,
  title={An entropic model of Gaia},
  author={Arthur, Rudy and Nicholson, Arwen},
  journal={Journal of theoretical biology},
  volume={430},
  pages={177--184},
  year={2017},
  publisher={Elsevier}
}

@article{arthur2022selection,
  title={Selection principles for Gaia},
  author={Arthur, Rudy and Nicholson, Arwen},
  journal={Journal of Theoretical Biology},
  volume={533},
  pages={110940},
  year={2022},
  publisher={Elsevier}
}

@article{arthur2023dice,
  title={Does Gaia Play Dice? Simple Models of Non-Darwinian Selection
},
  author={Arthur, Rudy and Nicholson, Arwen},
  journal={AstroBiology},
  volume={23},
  pages={1238–-1244},
  year={2023},
  publisher={Mary Ann Leibert Inc.}
}

@ARTICLE{nicholson2022,
       author = {{Nicholson}, A.~E. and {Daines}, S.~J. and {Mayne}, N.~J. and {Eager-Nash}, J.~K. and {Lenton}, T.~M. and {Kohary}, K.},
        title = "{Predicting biosignatures for nutrient-limited biospheres}",
      journal = {Monthly Notices of the Royal Astronomical Society},
         year = 2022,
        month = nov,
       volume = {517},
       number = {1},
        pages = {222-239},
          doi = {10.1093/mnras/stac2086},
archivePrefix = {arXiv},
       eprint = {2207.12961},
 primaryClass = {astro-ph.EP},
       adsurl = {https://ui.adsabs.harvard.edu/abs/2022MNRAS.517..222N}
}

@ARTICLE{quanz2022,
       author = {{Quanz}, S.~P. and {Ottiger}, M. and {Fontanet}, E. and {Kammerer}, J. and {Menti}, F. and {Dannert}, F. and {Gheorghe}, A. and {Absil}, O. and {Airapetian}, V.~S. and {Alei}, E. and {Allart}, R. and {Angerhausen}, D. and {Blumenthal}, S. and {Buchhave}, L.~A. and {Cabrera}, J. and {Carri{\'o}n-Gonz{\'a}lez}, {\'O}. and {Chauvin}, G. and {Danchi}, W.~C. and {Dandumont}, C. and {Defr{\'e}re}, D. and {Dorn}, C. and {Ehrenreich}, D. and {Ertel}, S. and {Fridlund}, M. and {Garc{\'\i}a Mu{\~n}oz}, A. and {Gasc{\'o}n}, C. and {Girard}, J.~H. and {Glauser}, A. and {Grenfell}, J.~L. and {Guidi}, G. and {Hagelberg}, J. and {Helled}, R. and {Ireland}, M.~J. and {Janson}, M. and {Kopparapu}, R.~K. and {Korth}, J. and {Kozakis}, T. and {Kraus}, S. and {L{\'e}ger}, A. and {Leedj{\"a}rv}, L. and {Lichtenberg}, T. and {Lillo-Box}, J. and {Linz}, H. and {Liseau}, R. and {Loicq}, J. and {Mahendra}, V. and {Malbet}, F. and {Mathew}, J. and {Mennesson}, B. and {Meyer}, M.~R. and {Mishra}, L. and {Molaverdikhani}, K. and {Noack}, L. and {Oza}, A.~V. and {Pall{\'e}}, E. and {Parviainen}, H. and {Quirrenbach}, A. and {Rauer}, H. and {Ribas}, I. and {Rice}, M. and {Romagnolo}, A. and {Rugheimer}, S. and {Schwieterman}, E.~W. and {Serabyn}, E. and {Sharma}, S. and {Stassun}, K.~G. and {Szul{\'a}gyi}, J. and {Wang}, H.~S. and {Wunderlich}, F. and {Wyatt}, M.~C. and {LIFE Collaboration}},
        title = "{Large Interferometer For Exoplanets (LIFE). I. Improved exoplanet detection yield estimates for a large mid-infrared space-interferometer mission}",
      journal = {Astronomy and Astrophysics},
         year = 2022,
        month = aug,
       volume = {664},
          eid = {A21},
        pages = {A21},
          doi = {10.1051/0004-6361/202140366},
archivePrefix = {arXiv},
       eprint = {2101.07500},
 primaryClass = {astro-ph.EP},
       adsurl = {https://ui.adsabs.harvard.edu/abs/2022A&A...664A..21Q}
}

@article{Seager_2013,
doi = {10.1088/0004-637X/775/2/104},
url = {https://dx.doi.org/10.1088/0004-637X/775/2/104},
year = {2013},
month = {sep},
publisher = {The American Astronomical Society},
volume = {775},
number = {2},
pages = {104},
author = {S. Seager and W. Bains and R. Hu},
title = {A BIOMASS-BASED MODEL TO ESTIMATE THE PLAUSIBILITY OF EXOPLANET BIOSIGNATURE GASES},
journal = {The Astrophysical Journal}
}

@article{lovelock1965physical,
  title={A physical basis for life detection experiments},
  author={Lovelock, James E},
  journal={Nature},
  volume={207},
  number={4997},
  pages={568--570},
  year={1965},
  publisher={Nature Publishing Group}
}

@article{Snellen:2021,
Author={Snellen, Ignas A. G. and  Snik, F. and  Kenworthy, M. and Albrecht, S. and  Anglada-Escudé, G. and  Baraffe, I. and  Baudoz, P. and Benz, W. and  Beuzit, J. L. and  Biller, B. and  Birkby, J. L. and  Boccaletti, A. and van Boekel, R. and  de Boer, J. and  Brogi, Matteo and  Buchhave, L. and Carone, L. and  Claire, M. and  Claudi, R. and  Demory, B. O. Désert, J. M. and  Desidera, S. and  Gaudi, B. S. and  Gratton, R. and Gillon, M. and  Grenfell, J. L. and  Guyon, O. and  Henning, T. and  Hinkley, S. and Huby, E. and  Janson, M. and  Helling, C. and  Heng, K. and  Kasper, M. and Keller, C. U. and  Krause, O. and  Kreidberg, L. and  Madhusudhan, N. and Lagrange, A. -M. and  Launhardt, R. and  Lenton, T. M. and Lopez-Puertas, M. and  Maire, A. -L. and  Mayne, N. and  Meadows, V. and Mennesson, B. and  Micela, G. and  Miguel, Y. and  Milli, J. and  Min, M. and de Mooij, E. and  Mouillet, D. and  N'Diaye, M. and  D'Orazi, V. and  Palle, E. and Pagano, I. and  Piotto, G. and  Queloz, D. and  Rauer, H. and  Ribas, I. and Ruane, G. and  Selsis, F. and Sozzetti, A. and  Stam, D. and  Stark, C. C. and Vigan, A. and  de Visser, Pieter},
Title={Detecting life outside our solar system with a large high-contrast-imaging mission},
Journal = {Experimental Astronomy},
Month = {October},
Year = {2021}
}

@article{kharecha2005coupled,
  title={A coupled atmosphere--ecosystem model of the early Archean Earth},
  author={Kharecha, P and Kasting, James and Siefert, J},
  journal={Geobiology},
  volume={3},
  number={2},
  pages={53--76},
  year={2005},
  publisher={Wiley Online Library}
}

@article{Krissansen-Totton:2022,
author = {Krissansen-Totton, Joshua  and
          Thompson, Maggie and
          Galloway, Max L. and 
          Fortney, Jonathan J.},
title = {Understanding planetary context to enable life
detection on exoplanets and test the Copernican
principle},
journal={Nature Astronomy},
volume={6(2)},
pages={189--198},
year={2022}
}

@article{Meadows:2018,
Author = {Meadows, Victoria S. and
Reinhard, Christopher T. and
Arney, Giada N. and
Parenteau , Mary N. and
Schwieterman, Edward W.  and
Domagal-Goldman, Shawn D.  and
Lincowski, Andrew P.  and
Stapelfeldt, Karl R.  and
Rauer, Heike  and
DasSarma, Shiladitya  and
Hegde, Siddharth  and
Narita, Norio  and
Deitrick, Russell  and
Lustig-Yaeger, Jacob  and
Lyons, Timothy W.  and
Siegler, Nicholas  and
Grenfell, J. Lee},
Title = {Exoplanet Biosignatures: Understanding Oxygen as a Biosignature in the Context of Its Environment},
Journal = {Astrobiology},
Year = {2018},
Volume = {18},
Issue = {6}}

@article{Catling:2018,
Author = {Catling, David C. and
Krissansen-Totton, Joshua and
Kiang, Nancy Y. and
Crisp, David and
Robinson, Tyler D. and
DasSarma, Shiladitya and
Rushby, Andrew J. and
Del Genio, Anthony and
Bains, William and
Domagal-Goldman, Shawn},
Title = {Exoplanet Biosignatures: A Framework for Their Assessment},
Journal = {Astrobiology},
Year = {2018},
Volume = {18:6}
}

@article{bruggeman:2014,
  title={A general framework for aquatic biogeochemical models},
  author={Bruggeman, Jorn and Bolding, Karsten},
  journal={Environmental modelling \& software},
  volume={61},
  pages={249--265},
  year={2014},
  publisher={Elsevier}
}

@article{Hunten:1973,
Author = {Hunten, D.M.},
Year = {1973},
Title = {The escape of light gases from planetary atmospheres},
Journal = {Journal of Atmospheric Science},
Volume = {30},
Pages = {1481--1494}}

@book{Walker:1977,
Author = {Walker, JCG},
Year = {1977},
Title = {Evolution of the Atmosphere Macmillan},
Publisher = {New York}}

@article{Boutle:2017,
Author = {Boutle, Ian A. and Mayne, Nathan J. and Drummond, Benjamin and Manners, James and Goyal, Jayesh and Lambert, Hugo F. and Acreman, David M. and Earnshaw, Paul D.},
Title = {Exploring the climate of Proxima B with the Met Office Unified Model},
Journal = {Astronomy \& Astrophysics},
Year = {2017},
Volume = {601},
Pages = {13}}

@ARTICLE{Eager:2020,
       author = {{Eager-Nash}, Jake K. and {Reichelt}, David J. and {Mayne}, Nathan J. and {Hugo Lambert}, F. and {Sergeev}, Denis E. and {Ridgway}, Robert J. and {Manners}, James and {Boutle}, Ian A. and {Lenton}, Timothy M. and {Kohary}, Krisztian},
        title = "{Implications of different stellar spectra for the climate of tidally locked Earth-like exoplanets}",
      journal = {Astronomy $\&$ Astrophysics},
         year = 2020,
        month = jul,
       volume = {639},
          eid = {A99},
        pages = {A99},
          doi = {10.1051/0004-6361/202038089},
archivePrefix = {arXiv},
       eprint = {2005.13002},
 primaryClass = {astro-ph.EP},
       adsurl = {https://ui.adsabs.harvard.edu/abs/2020A&A...639A..99E}
}

@article{Kral:1998,
  title={Hydrogen consumption by methanogens on the early Earth},
  author={Kral, Timothy A and Brink, Keith M and Miller, Stanley L and McKay, Christopher P},
  journal={Origins of Life and Evolution of the Biosphere},
  volume={28},
  number={3},
  pages={311--319},
  year={1998},
  publisher={Springer}
}

@article{Janssen:1996,
Author = {Janssen, Peter H. and Schuhmann, Alexandra and Bak, Friedhelm and Liesack, Werner}, 
Title = {Disproportionation of inorganic sulfur compounds by Desulfocapsa thiozymogenes gen. nov.}, 
Journal = {Arch. Microbiology},
Year = {1996},
Volume = {166},
Pages = {184--192}}

@article{Lynch:2019,
Author = {Lynch, T.A. and Wanga, Y.  and van Brunta, B. and Pachecob, D. and Janssenb,P.H.}, Title = {Modelling thermodynamic feedback on the metabolism of hydrogenotrophic methanogens},
Journal = {Journal of Theoretical Biology},
Volume = {477},
Year = {2019},
Pages = {14--23}}

@article{segura2005biosignatures,
  title={Biosignatures from Earth-like planets around M dwarfs},
  author={Segura, Ant{\'\i}gona and Kasting, James F and Meadows, Victoria and Cohen, Martin and Scalo, John and Crisp, David and Butler, Rebecca AH and Tinetti, Giovanna},
  journal={Astrobiology},
  volume={5},
  number={6},
  pages={706--725},
  year={2005},
  publisher={Mary Ann Liebert, Inc. 2 Madison Avenue Larchmont, NY 10538 USA}
}

@article{Berg1977,
  title={Physics of chemoreception},
  author={Berg, H.C. and Purcell, E.M.},
  journal={Biophysical Journal},
  volume={20},
  number={2},
  pages={193--219},
  year={1977},
  publisher={Elsevier}
}

@article{ARMSTRONG20081311,
title = {Nutrient uptake rate as a function of cell size and surface transporter density: A Michaelis-like approximation to the model of Pasciak and Gavis},
journal = {Deep Sea Research Part I: Oceanographic Research Papers},
volume = {55},
number = {10},
pages = {1311-1317},
year = {2008},
issn = {0967-0637},
doi = {https://doi.org/10.1016/j.dsr.2008.05.004},
url = {https://www.sciencedirect.com/science/article/pii/S0967063708000915},
author = {Robert A. Armstrong}
}

@article{pasciak1974transport,
  title={Transport limitation of nutrient uptake in phytoplankton 1},
  author={Pasciak, Walter J and Gavis, Jerome},
  journal={Limnology and Oceanography},
  volume={19},
  number={6},
  pages={881--888},
  year={1974},
  publisher={Wiley Online Library}
}

@article{pasciak1975transport,
  title={Transport limited nutrient uptake rates in Ditylum brightwellii 1},
  author={Pasciak, Walter J and Gavis, Jerome},
  journal={Limnology and Oceanography},
  volume={20},
  number={4},
  pages={604--617},
  year={1975},
  publisher={Wiley Online Library}
}

@article{madhusudhan2020interior,
  title={The interior and atmosphere of the habitable-zone exoplanet K2-18b},
  author={Madhusudhan, Nikku and Nixon, Matthew C and Welbanks, Luis and Piette, Anjali AA and Booth, Richard A},
  journal={The Astrophysical Journal},
  volume={891},
  number={1},
  pages={L7},
  year={2020},
  publisher={American Astronomical Society}
}

@article{shorttle2024distinguishing,
  title={Distinguishing oceans of water from magma on mini-Neptune K2-18b},
  author={Shorttle, Oliver and Jordan, Sean and Nicholls, Harrison and Lichtenberg, Tim and Bower, Dan J},
  journal={The Astrophysical Journal Letters},
  volume={962},
  number={1},
  pages={L8},
  year={2024},
  publisher={IOP Publishing}
}

@article{montet2015stellar,
  title={Stellar and planetary properties of K2 campaign 1 candidates and validation of 17 planets, including a planet receiving earth-like insolation},
  author={Montet, Benjamin T and Morton, Timothy D and Foreman-Mackey, Daniel and Johnson, John Asher and Hogg, David W and Bowler, Brendan P and Latham, David W and Bieryla, Allyson and Mann, Andrew W},
  journal={The Astrophysical Journal},
  volume={809},
  number={1},
  pages={25},
  year={2015},
  publisher={IOP Publishing}
}

@article{foreman2015systematic,
  title={A systematic search for transiting planets in the K2 data},
  author={Foreman-Mackey, Daniel and Montet, Benjamin T and Hogg, David W and Morton, Timothy D and Wang, Dun and Sch{\"o}lkopf, Bernhard},
  journal={The Astrophysical Journal},
  volume={806},
  number={2},
  pages={215},
  year={2015},
  publisher={IOP Publishing}
}

@article{tsiaras2019water,
  title={Water vapour in the atmosphere of the habitable-zone eight-Earth-mass planet K2-18 b},
  author={Tsiaras, Angelos and Waldmann, Ingo P and Tinetti, Giovanna and Tennyson, Jonathan and Yurchenko, Sergey N},
  journal={Nature Astronomy},
  volume={3},
  number={12},
  pages={1086--1091},
  year={2019},
  publisher={Nature Publishing Group UK London}
}

@article{cloutier2019confirmation,
  title={Confirmation of the radial velocity super-Earth K2-18c with HARPS and CARMENES},
  author={Cloutier, R and Astudillo-Defru, N and Doyon, R and Bonfils, Xavier and Almenara, J-M and Bouchy, F and Delfosse, X and Forveille, T and Lovis, C and Mayor, M and others},
  journal={Astronomy \& Astrophysics},
  volume={621},
  pages={A49},
  year={2019},
  publisher={EDP Sciences}
}

@article{sterzik2012biosignatures,
  title={Biosignatures as revealed by spectropolarimetry of Earthshine},
  author={Sterzik, Michael F and Bagnulo, Stefano and Palle, Enric},
  journal={Nature},
  volume={483},
  number={7387},
  pages={64--66},
  year={2012},
  publisher={Nature Publishing Group UK London}
}

@article{madhusudhan2021habitability,
  title={Habitability and biosignatures of hycean worlds},
  author={Madhusudhan, Nikku and Piette, Anjali AA and Constantinou, Savvas},
  journal={The Astrophysical Journal},
  volume={918},
  number={1},
  pages={1},
  year={2021},
  publisher={IOP Publishing}
}

@article{vitas2019towards,
  title={Towards a general definition of life},
  author={Vitas, Marko and Dobovi{\v{s}}ek, Andrej},
  journal={Origins of Life and Evolution of Biospheres},
  volume={49},
  number={1},
  pages={77--88},
  year={2019},
  publisher={Springer}
}

@article{liss1974flux,
  title={Flux of gases across the air-sea interface},
  author={Liss, Peter S and Slater, PG},
  journal={Nature},
  volume={247},
  number={5438},
  pages={181--184},
  year={1974},
  publisher={Nature Publishing Group UK London}
}

@article{zhang2018measurement,
  title={Measurement of thermal diffusivity for carbon dioxide (CO2) at T= 293.15--406.15 K and pressures up to 11 MPa by dynamic light scattering (DLS)},
  author={Zhang, Ying and Chen, Yutian and Zhan, Taotao and Zheng, Yu and Zheng, Xiong and He, Maogang},
  journal={Fluid Phase Equilibria},
  volume={474},
  pages={126--130},
  year={2018},
  publisher={Elsevier}
}

@article{ortega2024minimum,
  title={The minimum energy required to build a cell},
  author={Ortega-Arzola, Edwin and Higgins, Peter M and Cockell, Charles S},
  journal={Scientific Reports},
  volume={14},
  number={1},
  pages={5267},
  year={2024},
  publisher={Nature Publishing Group UK London}
}

@article{hutchison2016design,
  title={Design and synthesis of a minimal bacterial genome},
  author={Hutchison III, Clyde A and Chuang, Ray-Yuan and Noskov, Vladimir N and Assad-Garcia, Nacyra and Deerinck, Thomas J and Ellisman, Mark H and Gill, John and Kannan, Krishna and Karas, Bogumil J and Ma, Li and others},
  journal={Science},
  volume={351},
  number={6280},
  pages={aad6253},
  year={2016},
  publisher={American Association for the Advancement of Science}
}

@article{breuer2019essential,
  title={Essential metabolism for a minimal cell},
  author={Breuer, Marian and Earnest, Emmy E and Merryman, Chuck and Wise, Kim S and Sun, Lijie and Lynott, Michaela R and Hutchison, Clyde A and Smith, Hamilton O and Lapek, John D and Gonzalez, David J and others},
  journal={Elife},
  volume={8},
  pages={e36842},
  year={2019},
  publisher={eLife Sciences Publications, Ltd}
}

@article{eager20233d,
  title={3D climate simulations of the Archean find that methane has a strong cooling effect at high concentrations},
  author={Eager-Nash, Jake K and Mayne, Nathan J and Nicholson, Arwen E and Prins, Janke E and Young, Oakley CF and Daines, Stuart J and Sergeev, Denis E and Lambert, F Hugo and Manners, James and Boutle, Ian A and others},
  journal={Journal of Geophysical Research: Atmospheres},
  volume={128},
  number={6},
  pages={e2022JD037544},
  year={2023},
  publisher={Wiley Online Library}
}

@article{bar2018biomass,
  title={The biomass distribution on Earth},
  author={Bar-On, Yinon M and Phillips, Rob and Milo, Ron},
  journal={Proceedings of the National Academy of Sciences},
  volume={115},
  number={25},
  pages={6506--6511},
  year={2018},
  publisher={National Academy of Sciences}
}

@article{cavicchioli2019scientists,
  title={Scientists’ warning to humanity: microorganisms and climate change},
  author={Cavicchioli, Ricardo and Ripple, William J and Timmis, Kenneth N and Azam, Farooq and Bakken, Lars R and Baylis, Matthew and Behrenfeld, Michael J and Boetius, Antje and Boyd, Philip W and Classen, Aim{\'e}e T and others},
  journal={Nature Reviews Microbiology},
  volume={17},
  number={9},
  pages={569--586},
  year={2019},
  publisher={Nature Publishing Group UK London}
}

@article{west2005tectonic,
  title={Tectonic and climatic controls on silicate weathering},
  author={West, A Joshua and Galy, Albert and Bickle, Mike},
  journal={Earth and Planetary Science Letters},
  volume={235},
  number={1-2},
  pages={211--228},
  year={2005},
  publisher={Elsevier}
}

@article{brady1997seafloor,
  title={Seafloor weathering controls on atmospheric CO2 and global climate},
  author={Brady, Patrick V and G{\'\i}slason, Sigurdur R},
  journal={Geochimica et Cosmochimica Acta},
  volume={61},
  number={5},
  pages={965--973},
  year={1997},
  publisher={Elsevier}
}

@article{feist2006modeling,
  title={Modeling methanogenesis with a genome-scale metabolic reconstruction of Methanosarcina barkeri},
  author={Feist, Adam M and Scholten, Johannes CM and Palsson, Bernhard {\O} and Brockman, Fred J and Ideker, Trey},
  journal={Molecular systems biology},
  volume={2},
  number={1},
  pages={2006--0004},
  year={2006},
  publisher={John Wiley \& Sons, Ltd Chichester, UK}
}

@article{rowe2019methane,
  title={Methane-linked mechanisms of electron uptake from cathodes by Methanosarcina barkeri},
  author={Rowe, Annette R and Xu, Shuai and Gardel, Emily and Bose, Arpita and Girguis, Peter and Amend, Jan P and El-Naggar, Mohamed Y},
  journal={MBio},
  volume={10},
  number={2},
  pages={10--1128},
  year={2019},
  publisher={American Society for Microbiology 1752 N St., NW, Washington, DC}
}

@ARTICLE{arthur2025edge,
       author = {{Arthur}, R. and {Nicholson}, A.~E. and {Mayne}, N.~J.},
        title = "{Life on the Edge: Using Planetary Context to Enhance Biosignatures and Avoid False Positives}",
      journal = {\mnras},
         year = 2025,
        month = jul,
       volume = {541},
       number = {4},
        pages = {3664-3674},
          doi = {10.1093/mnras/staf1204},
archivePrefix = {arXiv},
       eprint = {2504.18431},
 primaryClass = {astro-ph.EP},
       adsurl = {https://ui.adsabs.harvard.edu/abs/2025MNRAS.541.3664A}
}

@article{leung2020energetic,
  title={Energetic basis of microbial growth and persistence in desert ecosystems},
  author={Leung, Pok Man and Bay, Sean K and Meier, Dimitri V and Chiri, Eleonora and Cowan, Don A and Gillor, Osnat and Woebken, Dagmar and Greening, Chris},
  journal={Msystems},
  volume={5},
  number={2},
  pages={10--1128},
  year={2020},
  publisher={American Society for Microbiology 1752 N St., NW, Washington, DC}
}

@article{mason2022microbial,
  title={Microbial storage and its implications for soil ecology},
  author={Mason-Jones, Kyle and Robinson, Serina L and Veen, GF and Manzoni, Stefano and van der Putten, Wim H},
  journal={The ISME Journal},
  volume={16},
  number={3},
  pages={617--629},
  year={2022},
  publisher={Oxford University Press}
}

@article{hobbie2013microbes,
  title={Microbes in nature are limited by carbon and energy: the starving-survival lifestyle in soil and consequences for estimating microbial rates},
  author={Hobbie, John E and Hobbie, Erik A},
  journal={Frontiers in microbiology},
  volume={4},
  pages={324},
  year={2013},
  publisher={Frontiers Media SA}
}

@article{wu2024microbes,
  title={Bacterial killing and the dimensions of bacterial death},
  author={Wu , Renfei and Li, Cong and Li, Jiuyi and Sjollema , Jelmer and Geertsema-Doornbusch, Gésinda I. and  de Haan-Visser, H. Willy and  Dijkstra , Emma S. C. and Ren , Yijin and Zhang , Zexin and Liu , Jian and Flemming, Hans C.  and  Busscher , Henk J.and van der Mei, Henny C.},
  journal={npj Biofilms Microbiomes },
  volume={10},
  pages={87},
  year={2024},
  publisher={Springer Nature}
}

@article{caro2025frozen,
author = {Caro, T. A. and McFarlin, J. M. and Maloney, A. E. and Jech, S. D. and Barker, A. J. and Douglas, T. A. and Barbato, R. A. and Kopf, S. H.},
title = {Microbial Resuscitation and Growth Rates in Deep Permafrost: Lipid Stable Isotope Probing Results From the Permafrost Research Tunnel in Fox, Alaska},
journal = {Journal of Geophysical Research: Biogeosciences},
volume = {130},
number = {9},
pages = {e2025JG008759},
doi = {https://doi.org/10.1029/2025JG008759},
url = {https://agupubs.onlinelibrary.wiley.com/doi/abs/10.1029/2025JG008759},
eprint = {https://agupubs.onlinelibrary.wiley.com/doi/pdf/10.1029/2025JG008759},
note = {e2025JG008759 2025JG008759},
year = {2025}
}

@article{servais1985,
  title={Rate of bacterial mortality in aquatic environments},
  author={Servais, Pierre and Billen, Gilles and Vives Rego, Jose},
  journal={American Society for Microbiology},
  volume={49},
  number={6},
pages = {1448-1454},
  year={1985}
}

@article{hill2023catalog,
  title={A catalog of habitable zone exoplanets},
  author={Hill, Michelle L and Bott, Kimberly and Dalba, Paul A and Fetherolf, Tara and Kane, Stephen R and Kopparapu, Ravi and Li, Zhexing and Ostberg, Colby},
  journal={The Astronomical Journal},
  volume={165},
  number={2},
  pages={34},
  year={2023},
  publisher={IOP Publishing}
}

@article{fujii2018exoplanet,
  title={Exoplanet biosignatures: observational prospects},
  author={Fujii, Yuka and Angerhausen, Daniel and Deitrick, Russell and Domagal-Goldman, Shawn and Grenfell, John Lee and Hori, Yasunori and Kane, Stephen R and Pall{\'e}, Enric and Rauer, Heike and Siegler, Nicholas and others},
  journal={Astrobiology},
  volume={18},
  number={6},
  pages={739--778},
  year={2018},
  publisher={Mary Ann Liebert, Inc. 140 Huguenot Street, 3rd Floor New Rochelle, NY 10801 USA}
}

@article{edmonds2022volcanic,
  title={Volcanic outgassing of volatile trace metals},
  author={Edmonds, Marie and Mason, Emily and Hogg, Olivia},
  journal={Annual Review of Earth and Planetary Sciences},
  volume={50},
  number={1},
  pages={79--98},
  year={2022},
  publisher={Annual Reviews}
}

@article{thompson2022case,
  title={The case and context for atmospheric methane as an exoplanet biosignature},
  author={Thompson, Maggie A and Krissansen-Totton, Joshua and Wogan, Nicholas and Telus, Myriam and Fortney, Jonathan J},
  journal={Proceedings of the National Academy of Sciences},
  volume={119},
  number={14},
  pages={e2117933119},
  year={2022},
  publisher={National Academy of Sciences}
}

@article{livingston2026young,
  title={A young progenitor for the most common planetary systems in the Galaxy},
  author={Livingston, John H and Petigura, Erik A and David, Trevor J and Masuda, Kento and Owen, James and Nesvorn{\`y}, David and Batygin, Konstantin and de Leon, Jerome and Mori, Mayuko and Ikuta, Kai and others},
  journal={Nature},
  volume={649},
  number={8096},
  pages={310--314},
  year={2026},
  publisher={Nature Publishing Group}
}

@article{zhu201830,
  title={About 30\% of Sun-like stars have Kepler-like planetary systems: a study of their intrinsic architecture},
  author={Zhu, Wei and Petrovich, Cristobal and Wu, Yanqin and Dong, Subo and Xie, Jiwei},
  journal={The Astrophysical Journal},
  volume={860},
  number={2},
  pages={101},
  year={2018},
  publisher={The American Astronomical Society}
}

@article{michal2018phymet2,
  title={PhyMet2: a database and toolkit for phylogenetic and metabolic analyses of methanogens},
  author={Micha{\l}, Burdukiewicz and Gagat, Przemys{\l}aw and Jab{\l}o{\'n}ski, S{\l}awomir and Chilimoniuk, Jaros{\l}aw and Gaworski, Micha{\l} and Mackiewicz, Pawe{\l} and Marcin, {\L}ukaszewicz},
  journal={Environmental microbiology reports},
  volume={10},
  number={3},
  pages={378--382},
  year={2018},
  publisher={Wiley Online Library}
}

@article{wang2023diffusion,
  title={Diffusion Coefficients of N2O and H2 in Water at Temperatures between 298.15 and 423.15 K with Pressures up to 30 MPa},
  author={Wang, Sijia and Zhou, Tong and Pan, Ziqing and Trusler, JP Martin},
  journal={Journal of Chemical \& Engineering Data},
  volume={68},
  number={6},
  pages={1313--1319},
  year={2023},
  publisher={ACS Publications}
}

@article{Weissman:2021,
author = {Weissman, Jake L.   and Hou, Shengwei and Fuhrman, Jed A. },
title = {Estimating maximal microbial growth rates from cultures, metagenomes, and single cells via codon usage patterns},
journal = {Proceedings of the National Academy of Sciences},
volume = {118},
number = {12},
pages = {e2016810118},
year = {2021},
doi = {10.1073/pnas.2016810118},

URL = {https://www.pnas.org/doi/abs/10.1073/pnas.2016810118},
eprint = {https://www.pnas.org/doi/pdf/10.1073/pnas.2016810118}
}

@article{miller2009atmospheric,
  title={The atmospheric signatures of super-Earths: how to distinguish between hydrogen-rich and hydrogen-poor atmospheres},
  author={Miller-Ricci, Eliza and Seager, Sara and Sasselov, Dimitar},
  journal={The Astrophysical Journal},
  volume={690},
  number={2},
  pages={1056--1067},
  year={2009},
  publisher={The American Astronomical Society}
}

@article{hunten1976hydrogen,
  title={Hydrogen loss from the terrestrial planets},
  author={Hunten, DM and Donahue, Thomas M},
  journal={In: Annual review of earth and planetary sciences. Volume 4.(A76-37259 18-91) Palo Alto, Calif., Annual Reviews, Inc., 1976, p. 265-292.},
  volume={4},
  pages={265--292},
  year={1976}
}

@article{savanov2021activity,
  title={The Activity of Stars with Planetary Systems and Its Impact on the Loss of Atmosphere by Hot Exoplanets},
  author={Savanov, IS and Shematovich, VI},
  journal={Astrophysical Bulletin},
  volume={76},
  number={4},
  pages={450--471},
  year={2021},
  publisher={Springer}
}

@article{madhusudhan2023chemical,
  title={Chemical conditions on Hycean worlds},
  author={Madhusudhan, Nikku and Moses, Julianne I and Rigby, Frances and Barrier, Edouard},
  journal={Faraday Discussions},
  volume={245},
  pages={80--111},
  year={2023},
  publisher={Royal Society of Chemistry}
}

@misc{lunine2025characterization,
  title={Characterization of exoplanets in the James Webb Space Telescope era},
  author={Lunine, Jonathan I and Bahcall, Neta},
  journal={Proceedings of the National Academy of Sciences},
  volume={122},
  number={39},
  pages={e2507109122},
  year={2025},
  publisher={National Academy of Sciences}
}

@article{bezard2022methane,
  title={Methane as a dominant absorber in the habitable-zone sub-Neptune K2-18 b},
  author={B{\'e}zard, Bruno and Charnay, Benjamin and Blain, Doriann},
  journal={Nature Astronomy},
  volume={6},
  number={5},
  pages={537--540},
  year={2022},
  publisher={Nature Publishing Group UK London}
}

@article{kiang2018exoplanet,
  title={Exoplanet biosignatures: at the dawn of a new era of planetary observations},
  author={Kiang, Nancy Y and Domagal-Goldman, Shawn and Parenteau, Mary N and Catling, David C and Fujii, Yuka and Meadows, Victoria S and Schwieterman, Edward W and Walker, Sara I},
  journal={Astrobiology},
  volume={18},
  number={6},
  pages={619--629},
  year={2018},
  publisher={SAGE Publications Sage CA: Los Angeles, CA}
}

@article{wogan2020abundant,
  title={Abundant atmospheric methane from volcanism on terrestrial planets is unlikely and strengthens the case for methane as a biosignature},
  author={Wogan, Nicholas and Krissansen-Totton, Joshua and Catling, David C},
  journal={The Planetary Science Journal},
  volume={1},
  number={3},
  pages={58},
  year={2020},
  publisher={The American Astronomical Society}
}

@book{broecker1982tracers,
  title={Tracers in the Sea},
  author={Broecker, W.S. and Beng, Z. and Lamont-Doherty Geological Observatory},
  series={A Publication of the Lamont-Doherty geological observator},
  url={https://books.google.co.uk/books?id=uFq20AEACAAJ},
  year={1982},
  publisher={Lamont-Doherty Geological Observatory, Columbia University}
}

@article{pavlova2022bacterial,
  title={Bacterial cell shape: Some features of ultrastructure, evolution, and ecology},
  author={Pavlova, MD and Asaturova, AM and Kozitsyn, AE},
  journal={Biology Bulletin Reviews},
  volume={12},
  number={3},
  pages={254--265},
  year={2022},
  publisher={Springer}
}

@article{moger2023evolution,
  title={Evolution of a minimal cell},
  author={Moger-Reischer, Roy Z and Glass, John I and Wise, Kenneth S and Sun, Lijie and Bittencourt, Debora MC and Lehmkuhl, Blake K and Schoolmaster Jr, DR and Lynch, Michael and Lennon, John T},
  journal={Nature},
  volume={620},
  number={7972},
  pages={122--127},
  year={2023},
  publisher={Nature Publishing Group UK London}
}

@article{pahlow1997impact,
  title={Impact of cell shape and chain formation on nutrient acquisition by marine diatoms},
  author={Pahlow, Markus and Riebesell, Ulf and Wolf-Gladrow, Dieter A},
  journal={Limnology and Oceanography},
  volume={42},
  number={8},
  pages={1660--1672},
  year={1997},
  publisher={Wiley Online Library}
}

@article{karp1996nutrient,
  title={Nutrient fluxes to planktonic osmotrophs in the presence of fluid motion},
  author={Karp-Boss, L and Boss, E and Jumars, PA and others},
  journal={Oceanography and marine biology},
  volume={34},
  pages={71--108},
  year={1996},
  publisher={UCL PRESS LIMITED}
}

@article{fernandez2026atmospheric,
  title={The atmospheric composition of the sub-Neptune K2-18, b and insights into its formation},
  author={Fern{\'a}ndez-Rodr{\'\i}guez, Gareb and Morello, Giuseppe and Tan, Jonathan C and Pall{\'e}, Enric and Swain, Mark R and Poultourtzidis, Efthymios and Biagini, Alfredo and Changeat, Quentin and Jiang, Chengzi and Pozuelos, Francisco J and others},
  journal={Astronomy \& Astrophysics},
  year={2026},
  publisher={EDP Sciences}
}

@article{madhusudhan2023carbon,
  title={Carbon-bearing molecules in a possible Hycean atmosphere},
  author={Madhusudhan, Nikku and Sarkar, Subhajit and Constantinou, Savvas and Holmberg, M{\aa}ns and Piette, Anjali AA and Moses, Julianne I},
  journal={The Astrophysical Journal Letters},
  volume={956},
  number={1},
  pages={L13},
  year={2023},
  publisher={IOP Publishing}
}

@article{benneke2024jwst,
  title={JWST Reveals CH $ \_4 $, CO $ \_2 $, and H $ \_2 $ O in a Metal-rich Miscible Atmosphere on a Two-Earth-Radius Exoplanet},
  author={Benneke, Bj{\"o}rn and Roy, Pierre-Alexis and Coulombe, Louis-Philippe and Radica, Michael and Piaulet, Caroline and Ahrer, Eva-Maria and Pierrehumbert, Raymond and Krissansen-Totton, Joshua and Schlichting, Hilke E and Hu, Renyu and others},
  journal={arXiv preprint arXiv:2403.03325},
  year={2024}
}

@article{holmberg2024possible,
  title={Possible Hycean conditions in the sub-Neptune TOI-270 d},
  author={Holmberg, M{\aa}ns and Madhusudhan, Nikku},
  journal={Astronomy \& Astrophysics},
  volume={683},
  pages={L2},
  year={2024},
  publisher={EDP Sciences}
}

@article{bell2023methane,
  title={Methane throughout the atmosphere of the warm exoplanet WASP-80b},
  author={Bell, Taylor J and Welbanks, Luis and Schlawin, Everett and Line, Michael R and Fortney, Jonathan J and Greene, Thomas P and Ohno, Kazumasa and Parmentier, Vivien and Rauscher, Emily and Beatty, Thomas G and others},
  journal={Nature},
  volume={623},
  number={7988},
  pages={709--712},
  year={2023},
  publisher={Nature Publishing Group UK London}
}

@article{roy2024jwst,
  title={JWST reveals abundant methane and depleted carbon dioxide on the temperate sub-Neptune LP791-18c},
  author={Roy, Pierre-Alexis and Benneke, Bj{\"o}rn and Piaulet, Caroline and Coulombe, Louis-Philippe and Fournier-Tondreau, Marylou and Lafreniere, David},
  journal={AAS/Division for Extreme Solar Systems Abstracts},
  volume={56},
  number={4},
  pages={502--04},
  year={2024}
}

@article{trappist2024roadmap,
  title={A roadmap for the atmospheric characterization of terrestrial exoplanets with JWST},
  author={{TRAPPIST-1 JWST Community Initiative}},
  journal={Nature Astronomy},
  volume={8},
  number={7},
  pages={810--818},
  year={2024},
  publisher={Nature Publishing Group UK London}
}

@article{seager2025prospects,
  title={Prospects for detecting signs of life on exoplanets in the JWST era},
  author={Seager, Sara and Welbanks, Luis and Ellerbroek, Lucas and Bains, William and Petkowski, Janusz J},
  journal={Proceedings of the National Academy of Sciences},
  volume={122},
  number={39},
  pages={e2416188122},
  year={2025},
  publisher={National Academy of Sciences}
}




\appendix

\onecolumn
\section{Additional material}

In this Appendix we include further details of the model, including results justifying the model assumptions made and including additional data to support our conclusions. The section order follows the order in which they are referenced in the main article. 

\subsection{Ocean-atmosphere gas exchange}
\label{section:appendix_gas_exchange}

Section \ref{subsection:ocean_atmo_gas_exchange} briefly explains the exchange of gas between the atmosphere and ocean in our model.
The rate of exchange of gases between the ocean and the atmosphere is calculated following the stagnant layer boundary model \citep{liss1974flux}. For any chemical the rate of exchange of molecules between the ocean and atmosphere is calculated from the relative concentration of the gas in the ocean and in the atmosphere, given by
\begin{equation}
    \Phi_{X} = v_{p}\ . \ (\alpha(X)\ .\ pX-[X]_aq),
\end{equation}
where $\Phi_{x}$ is the flux of chemical $X$ from the atmosphere into the ocean (here, given in terms of moles), $v_{p}$ is the piston velocity of the chemical $X$, $\alpha(X)$ is the solubility of chemical $X$ (the Henry's law coefficient), $pX$ is the partial pressure of $X$ (in bars) and $[X]_{aq}$ is the dissolved concentration of $X$ in $mol/m^{3}$. $v_{p}$ is calculated by dividing the diffusivity of $X$ by the thickness of the stagnant layer film which we assume to be $z_{film} = 40\ \mu m$ following \citet{kharecha2005coupled}. Table \ref{table:gas_air_exchange} shows the values used when calculating the ocean-atmosphere exchange of gases in the model. The value of $z_{film}$ used here is standard in models of atmosphere-ocean gas exchange on Earth, and is derived from experimental data \citep{broecker1982tracers, kharecha2005coupled}. Investigating how this parameter would differ under different planetary contexts (e.g. larger planets) we leave for future work.

\begin{table}
    \centering
    \begin{tabular}{c|c|c|c}
     \hline \hline
        Chemical & Piston velocity ($m\ s^{-1}$) & Diffusivity ($m^{2}s^{-1}$) & Solubility ($mol\ L^{-1}bar^{-1}$) \\
          \hline \hline 
          & & &\\
      $CO_{2}$ & $6.7 \times 10^{-4}$ & $2.67\times 10^{-8}\ ^{\alpha}$ & $0.035\ ^{\beta}$ \\
      $H_{2}$  & $1.3 \times 10^{-4}\ ^{*}$ & $5.0\times 10^{-9}\ ^{*}$ & $7.8\times 10^{-4}\ ^{*}$ \\
      $CH_{4}$ & $4.5\times 10^{-5}\ ^{*}$ &  $1.8\times 10^{-9}\ ^{*}$ & $1.4\times 10^{-3}\ ^{*}$ \\
    \end{tabular}
    \caption{
        Parameter values for air-ocean exchange of $CO_{2}$, $H_{2}$, and $CH_{4}$. Values marked with a $*$ are taken from \protect\cite{kharecha2005coupled}, $\alpha$ from \protect\cite{zhang2018measurement}, and $\beta$ from \href{https://webbook.nist.gov/chemistry/}{https://webbook.nist.gov/chemistry/}. Piston velocities are calculated assuming a stagnant boundary layer thickness of $z_{film} = 40\mu m$. We assume a fixed temperature of $25^{o}C$ when calculating the gas exchange between the atmosphere and ocean.
    }
    \label{table:gas_air_exchange}
\end{table}

\subsection{Removing the ATP  maintenance cost of a cell}
\label{section:appendix_A0}

As discussed in Section \ref{subsection:microbe_model} we remove the ATP cell maintenance cost and, instead of varying this parameter as was done in \citet{nicholson2022}, we vary the energetic cost of biosynthesis. Here we show that a changing ATP maintenance cost, or a changing biosynthesis cost affect the biosignature in the same way and thus are equivalent ways of representing the biosphere for our purposes. Using a model based on \citet{nicholson2022} (see paper for full model details) and a default ATP maintenance cost value of $c_{0}\ =\ 1.75 \times 10^{-3}\ mol_{ATP} g^{-1} h^{-1}$ \citep{Lynch:2019} we vary the ATP cost using the values $0.0$,  $0.1 c_{0}$, $0.5 c_{0}$, $c_{0}$, $1.5 c_{0}$ and $2.0 c_{0}$. We repeat these experiments for energetic costs of generating 1 mole of ATP of $32.5\ kJ\ mol_{ATP}$ (the result under standard conditions) and $75\ kJ\ mol_{ATP}$ \citep[the average cost for Methanosarcina barkeri][]{Lynch:2019}. We then compare to the model outlined in the main body of this paper where there is no ATP cost but instead the energy cost of synthesising biomass is varied.

   \begin{figure*}
        \centering
        \includegraphics[width=0.5\hsize]{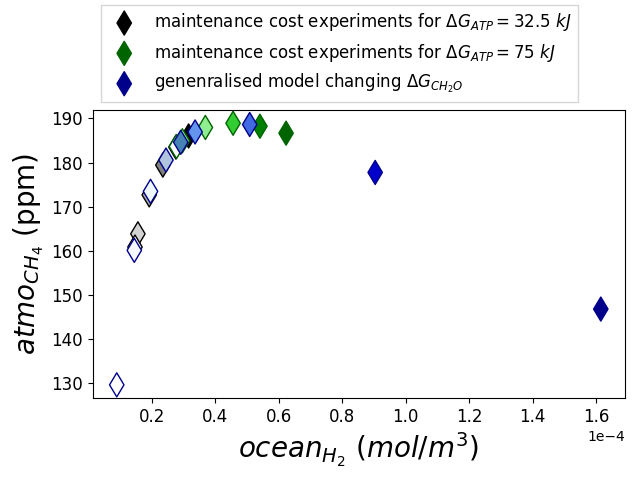}
        \caption{Panel b) shows a comparison between the original model $ATP$ maintenance cost experiments for both cases where $\Delta G_{ATP} = 32.5\ kJ$ (shown in black), and $\Delta G_{ATP} = 75\ kJ$ (shown in green), and the generalised cell model where $\Delta G_{CH_{2}O}$ is varied (data shown in blue). For all data points the default values for cell size, cell protein density and cell death rate are those listed in Table \ref{table:microbes}.}
        \label{fig:ATP_removal}%
    \end{figure*}

Figure \ref{fig:ATP_removal} shows a comparison of $atmo_{CH_{4}}$ as a function of $ocean_{H_{2}}$ for the maintenance cost experiments using the original model for both $\Delta G_{ATP}=32.5\ kJ$ and $\Delta G_{ATP}=75\ kJ$ and for the model outlined in the main text (referred to as the `generalised model') where the the cost of biomass synthesis $\Delta G_{CH_{2}O}$ is varied instead. The data in Figure \ref{fig:ATP_removal} follows the same curve showing that changing $\Delta G_{ATP}$ or $\Delta G_{CH_{2}O}$ have equivalent effects on the biosphere. 

\subsection{Calculating the default biological parameters in Section \ref{subsection:microbe_model}}
\label{section:appendix_A1}

We adopted the values for Methanosarcina barkeri from \citep{Lynch:2019} for use in our model. The maintenance requirement of a microbe was reported as $2.6 \times 10^{-19} mol_{ATP}\ cell^{-1}\ s^{-1}$. As we vary the size of our cells in this work we calculated the maintenance cost per gram of biomass. Methanosarcina barkeri cells have been measured as being roughly $1-2\mu m$ in length \citep{rowe2019methane} and so we use a radius of $1\mu m$ as our default for the cell size. An ATP maintenance cost for Methanosarcina barkeri cells has been measured to be $1.75 \time 10^{-3} mol_{ATP}\ g^{-1}_{cell}\ h^{-1}$ \citep{feist2006modeling} and we use this value as our default in this work.  We calculate our default value for the $CH_{2}O$ content of a cell per volume, by assuming a dry cell mass of $4.44\times10^{-13}g$ \citep{Janssen:1996} and make the simplification that a cell consists only of $CH_{2}O$, which has a molecular mass of $30.031 g\ mol^{-1}$. We assume a perfectly spherical cell of radius $10^{-6}m$ and this yields a $CH_{2}O$ density of $\approx 3530 mol_{CH_{2}O}\ m^{-3}$. \citet{Lynch:2019} use a value of $2.36\times10^{13}cell\ mol^{-1}_{ATP}$ for the number of cells that can be made using 1 mole of ATP. Again using the assumption of a cell mass of $4.44\times10^{-13}g$ \citep{Janssen:1996} and the simplification of a cell consisting entirely of $CH_{2}O$ leads to a cost of building 1 mole of $CH_{2}O$ being $\approx 3 mol_{ATP}$. In our model we assume a constant rate of removal of cells from the population due to cell death. This varies greatly between species of microbes and depends strongly on the environment. We choose to remove $2\%$ of the population per hour as our default value and vary this significantly to capture the sensitivity of the model to this parameter.

\subsection{Model time steps}
\label{section:appendix_A3}

As described in Section \ref{subsection:model_structure} the model is initially run abiotically for $20,000$ years before seeding with life to allow the atmosphere and ocean to reach equilibrium. Life is then seeded at $t = 20,000$ years and the model is run for a further $20,000$ years.

The model iteration is described by the following sequence.
\begin{enumerate}
\item update atmospheric $H_{2}$, $CO_{2}$, and $CH_{4}$ due to inflows from abiotic sources, and outflows due to the processes outlined in Sections \ref{section:methods}
    \item calculate the transfer of $H_{2}$, $CO_{2}$, and $CH_{4}$ between the atmosphere and the ocean
    \item for $365 \times 60$ steps, each representing an hour:
    \begin{enumerate}
    
    \item calculate the energy obtained from 1$mol$ of $H_{2}$ as per Equation \ref{eq:gibbs}
    \item calculate $H_{2}$ available to a cell as per Equation \ref{eq:diff_limited} multiplied by current population of the biosphere
    \item mol of $H_{2}$ required for energy generation per mol of $H_{2}$ converted to biomass is given by Equation \ref{eq:ratio}
    \item assign available $H_{2}$ to energy generation or biomass as per ratio
    \item add created biomass to the biomass of biosphere
    \item amount of $CH_{4}$ generated by biosphere is $4\times H_{2}$ used for energy
    \item update the ocean concentrations of $H_{2}$, $CO_{2}$, and $CH_{4}$ due to biological activity
\end{enumerate}
\end{enumerate}

\subsection{The effect of methane recycling on the model dynamics}
\label{section:Appendix_A4}

In Section \ref{subsection:changing_ch2o_cost} we showed that two different levels of atmospheric methane were possible for the same influx of $H_{2}$ to the ocean. This is due to the treatment of $CH_{4}$ and $H_{2}$ in the atmosphere.

Methane recycling in the atmosphere, as described in Section \ref{section:methods} and the dependence of $H_{2}$ loss from the top of the atmosphere on both the quantity of $H_{2}$ and $CH_{4}$ in the atmosphere lead to a non-linear relationship between the level of $H_{2}$ in the atmosphere and the concentration of $H_{2}$ in the ocean. The effect of methane recycling on the  model dynamics is larger than that of $H_{2}$ loss, with the recycling contributing to the peak in $atmo_{CH_{4}}$ seen in Figure \ref{fig:3}. This is as effectively adds an additional source of $H_{2}$ to the atmosphere on top of the abiotic influx.

\begin{figure*}
\centering
\begin{subfigure}{.45\textwidth}
  \centering
  \includegraphics[scale=0.47]{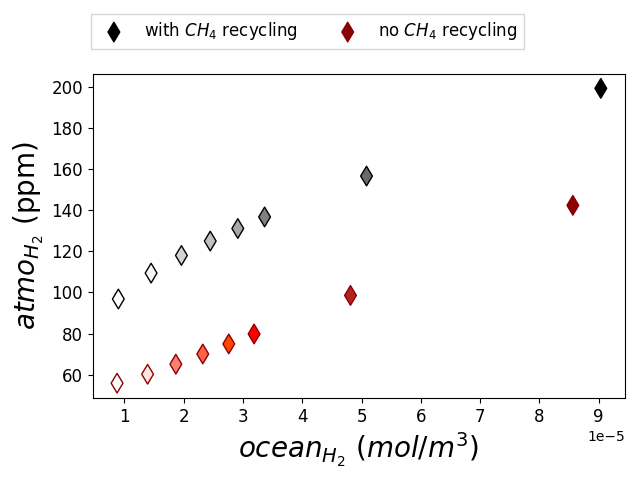}
  \caption{Level of atmospheric $H_{2}$ as a function of the concentration of $H_{2}$ in the ocean.}
  \label{fig:5_a}
\end{subfigure}%
\hfill
\begin{subfigure}{.45\textwidth}
  \centering
  \includegraphics[scale=0.47]{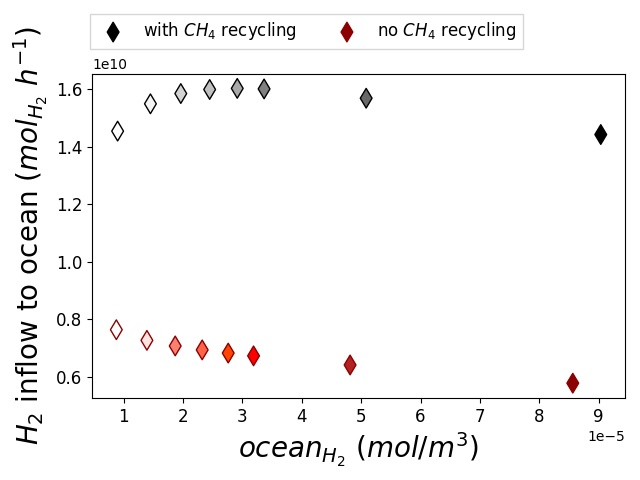}
  \caption{Rate of $H_{2}$ inflow into the ocean as a function of the concentration of $H_{2}$ in the ocean.}
  \label{fig:5_b}
\end{subfigure}

\centering
\begin{subfigure}{.45\textwidth}
  \centering
  \includegraphics[scale=0.47]{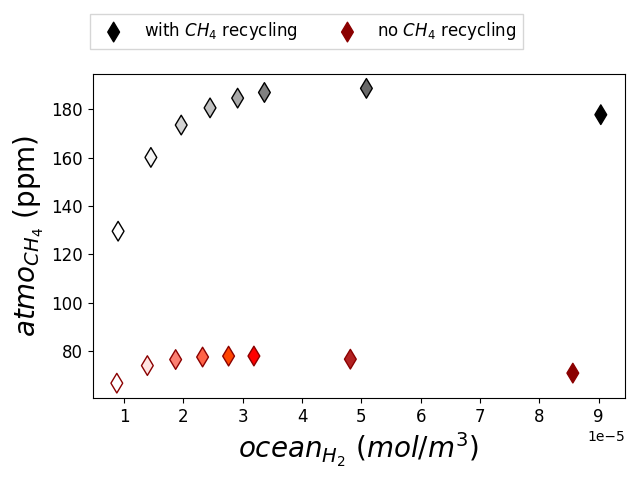}
  \caption{Level of atmospheric $CH_{4}$ as a function of the concentration of $H_{2}$ in the ocean.}
  \label{fig:5_c}
\end{subfigure}
\hfill
\begin{subfigure}{.45\textwidth}
  \centering
  \includegraphics[scale=0.47]{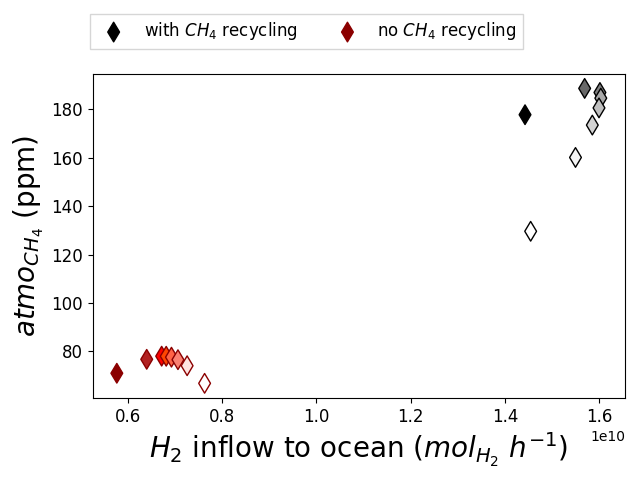}
  \caption{Level of atmospheric $CH_{4}$ as a function of the rate of $H_{2}$ inflow into to the ocean.}
  \label{fig:5_d}
\end{subfigure}

\centering
\caption{Panels showing data from experiments with differing $CH_{2}O$ synthesis costs both with methane recycling as described in Section \ref{section:methods}, shown in black, and without methane recycling, shown in red, where instead the same quantity of methane is removed from the atmosphere each timestep. For all data $\Delta G_{CH_{4}}$ is given by Equation \ref{eq:gibbs}. Marker colour saturation indicates the parameter value for $CH_{2}O$ synthesis cost with a darker marker indicating a higher $CH_{2}O$ cost.}
\label{fig:5}
\end{figure*}

Figure \ref{fig:5_a} shows the level of $H_{2}$ in the atmosphere as a function of the concentration of $H_{2}$ in the ocean for scenarios both with and without $CH_{4}$ recycling. For the experiments without $CH_{4}$ recycling (the red data in Figure \ref{fig:5}) the same amount of $CH_{4}$ is removed each timestep, however instead of being recycled into $H_{2}$ it is simply removed from the atmosphere. Figure \ref{fig:5_a} shows that with $CH_{4}$ recycling the relationship between $atmo_{H_{2}}$ and $ocean_{H_{2}}$ is non-linear. Figure \ref{fig:5_b} shows the same data but for the rate of $H_{2}$ inflow to the ocean against $ocean_{H_{2}}$. Without $CH_{4}$ recycling the $H_{2}$ inflow decreases as $ocean_{H_{2}}$ increases, however with $CH_{4}$ recycling the flow of $H_{2}$ to the ocean initially increases before decreasing after $ocean_{H_{2}} \approx 3\times 10^{-5} mol/m^{3}$. The peak in $atmo_{CH_{4}}$ is not solely down to this behaviour and occurs in the absence of methane recycling. Figure \ref{fig:5_c} shows $atmo_{CH_{4}}$ as a function of $ocean_{H_{2}}$ for experiments with and without methane recycling and a peak in $atmo_{CH_{4}}$ is seen in both, although $atmo_{CH_{4}}$ peaks earlier without recycling at $ocean_{H_{2}}\approx 3\times 10^{-5} mol/m^{3}$ as a function of $ocean_{H_{2}}\approx 5\times 10^{-5} mol/m^{3}$ with methane recycling. Figure \ref{fig:5_d} shows $atmo_{CH_{4}}$ as a function of $H_{2}$ inflow to the ocean for experiments both with and without methane recycling and shows that without methane recycling only one $atmo_{CH_{4}}$ scenario is possible for each $H_{2}$ inflow rate.


\bsp	
\label{lastpage}
\end{document}